\documentclass[journal]{IEEEtran}

\usepackage{cite,latexsym,amssymb,amsmath,algorithm,setspace}
\usepackage[dvips]{graphicx}
\usepackage[caption=false,font=footnotesize]{subfig}

\newtheorem{lemma}{Lemma}

\newtheorem{remark}{Remark}

\newtheorem{proposition}{Proposition}

\newenvironment{Proof}[1]{\medskip\par\noindent{\bf Proof:\,}\,#1}{{\mbox{\,$\blacksquare$}\medskip\par}}

\newcommand{\naive}{na\"{\i}ve~}

\makeatletter

\newcommand{\bb}{\mathbf}
\allowdisplaybreaks[1]

\begin{document}

   \author{
   \thanks{This work was supported by NSF Grants CNS 09-64364 and CCF 14-22347. This work was presented in part at the International Conference on Communications, Workshop on Green Broadband Access: Energy Efficient Wireless and Wired Network Solutions, June 2013 \cite{tutuncuoglu2013optimum_twrc}.}
	\thanks{The authors are with the department of Electrical Engineering, Pennsylvania State University, University Park, PA 16802 USA (e-mail: kaya@psu.edu; varan@psu.edu; yener@ee.psu.edu).}
	Kaya Tutuncuoglu, Burak Varan, and Aylin Yener
	}

   \title{Throughput Maximization for Two-way Relay Channels with Energy Harvesting Nodes: \\ The Impact of Relaying Strategies}

\maketitle

\begin{abstract}
In this paper, we study the two-way relay channel with energy harvesting nodes. In particular, we find transmission policies that maximize the sum-throughput for two-way relay channels when the relay does not employ a data buffer. The relay can perform decode-and-forward, compress-and-forward, compute-and-forward or amplify-and-forward relaying. Furthermore, we consider throughput improvement by dynamically choosing relaying strategies, resulting in hybrid relaying strategies. We show that an iterative generalized directional water-filling algorithm solves the offline throughput maximization problem, with the achievable sum-rate from an individual or hybrid relaying scheme. In addition to the optimum offline policy, we obtain the optimum online policy via dynamic programming. We provide numerical results for each relaying scheme to support the analytic findings, pointing out to the advantage of adapting the instantaneous relaying strategy to the available harvested energy.
\end{abstract}


\section{Introduction}
\label{sect_intro}

Wireless networks consisting of energy harvesting nodes continue to gain significance in the area of green communications \cite{niyato2007sleep, lei2009generic, yang2012optimal, tutuncuoglu2012optimum, ozel2011fading, ho2012optimal, devillers:imperfections, tutuncuoglu2014inefficient, luo2013optimal, tutuncuoglu2012ita, mahdavi2013energy, yang2012mac, yang2012broadcasting, ozel2012optimalbc, tutuncuoglu2012sum, antepli2011optimal, blasco2013low, gunduz2011two, orhan2012optimal, orhan2012energy, orhan2013throughput, ahmed2012power, luo2012optimal2, huang2013throughput, ahmed2013joint}. These networks {\it harvest} energy from external sources in an intermittent fashion, and consequently require careful management of the available energy. 

There is considerable recent research on energy management for energy harvesting networks. Reference \cite{yang2012optimal} considers an energy harvesting transmitter with energy and data arrivals, and an infinite size battery to store the harvested energy, and shows the optimality of a piecewise constant power policy for minimization of completion time of a file transfer. In \cite{tutuncuoglu2012optimum}, the throughput maximization problem is solved when the energy storage capacity of the battery is limited. It is shown that the transmission power policy is again piecewise constant, changing only when the battery is full or depleted. Extension of the model in \cite{tutuncuoglu2012optimum} to fading channels is studied in \cite{ozel2011fading} where a directional water-filling algorithm is shown to yield the optimum transmission policy. Reference \cite{ho2012optimal} also considers throughput maximization for a fading channel under the same assumption. The impact of degradation and imperfections of energy storage on the throughput maximizing policies is studied in \cite{devillers:imperfections, tutuncuoglu2014inefficient, luo2013optimal}. The single user channel with an energy harvesting transmitter and an energy harvesting receiver is considered in \cite{tutuncuoglu2012ita}, and decoding and sampling strategies for energy harvesting receivers is considered in \cite{mahdavi2013energy}. Various multi-user energy harvesting networks have also been studied to date; including multiple access, broadcast, and interference channels with energy harvesting nodes \cite{yang2012mac, yang2012broadcasting, ozel2012optimalbc, tutuncuoglu2012sum, antepli2011optimal, blasco2013low}. In addition to these multi-user setups, variations of the energy harvesting relay channel are studied in \cite{gunduz2011two, orhan2012optimal, orhan2012energy, orhan2013throughput, ahmed2012power, luo2012optimal2, huang2013throughput}, including multiple energy harvesting relays \cite{ahmed2013joint}.

In this work, we study the simplest network setup that embodies a cooperative communication scenario with two-directional information flow, with the goal of identifying design insights unique to such scenarios. This leads to the investigation of bi-directional communication with energy harvesting nodes. Specifically, we study the so-called separated two-way relay channel\footnote{Usually referred to as the two-way relay channel, as we will in the sequel.} with energy harvesting nodes. The channel is separated in the sense that the users cannot hear each other directly, i.e., communication is only possible through the relay. This model is relevant and of interest for peer-to-peer communications, or for any scenario where a pair of nodes exchange information, and avails the relay node to implement strategies to convey both messages simultaneously. The two-way relay channel (TWRC) with conventional (non-energy-harvesting) nodes is studied with various relaying strategies such as amplify-and-forward, decode-and-forward, compress-and-forward \cite{kim2008performance, kim2011achievable, rankov2006achievable}, and compute-and-forward\cite{nam2010capacity} in half-duplex \cite{kim2008performance, kim2011achievable} and full-duplex \cite{gunduz2008rate, rankov2006achievable} models. It is observed that different relaying schemes outperform the others for different ranges of transmit powers.

In this paper, we identify transmission power policies for the energy harvesting two-way relay channel (EH-TWRC) which maximize the sum-throughput. The energy harvesting relay can perform amplify-and-forward, decode-and-forward, compress-and-forward, or compute-and-forward relaying. Due to intermittent energy availability, the channel calls for relaying strategies that adapt to varying transmit powers. For this purpose, we introduce a relay that can dynamically change its relaying strategy, resulting in what we term hybrid relaying strategies. We derive the properties of the optimal offline transmission policy, where energy arrivals are known non-causally, with the goal of gaining insights into its structure. Next, we show that an iterative generalized directional water-filling algorithm solves the sum-throughput maximization problem for all relaying strategies. We next find the optimal online transmission policy by formulating and solving a dynamic program, where the energy states of the nodes are known causally. We compute optimal policies for different relaying strategies and provide numerical comparisons of their sum-throughputs. Our contribution includes generalization of directional water-filling \cite{ozel2011fading} to an interactive communication scenario with multiple energy harvesting terminals in the offline setting, as well as the identification of optimal policies in the online setting. The interactive communication scenario considered in this paper is the catalyst that can drastically change the resulting power allocation algorithms in the energy harvesting setting. The two-way relay channel is the simplest multi terminal network model that demonstrates this interaction, and hence is the model considered. We observe that the relaying strategy has a significant impact on the optimum transmission policy, i.e., transmit powers and phase durations, and that hybrid relaying can provide a notable throughput improvement for the EH-TWRC.

The remainder of the paper is organized as follows. The system model is described in Section~\ref{sect_model}. In Section~\ref{sect_hybrid}, a hybrid relaying scheme where the relay can alter its strategy depending on the instantaneous powers is introduced. In Section~\ref{sect_properties}, the sum-throughput maximization problem is presented for an EH-TWRC, and is divided into subproblems that can be solved separately. In Section~\ref{sect_identify}, the iterative generalized directional water-filling algorithm is proposed to find an optimal policy for the EH-TWRC. The online policy based on dynamic programming is provided in Section~\ref{sect_online}. Numerical results are presented in Section~\ref{sect_numerical}. The paper is concluded in Section~\ref{sect_conclusion}.

\begin{figure}[t]
\centering
\includegraphics[width=\linewidth]{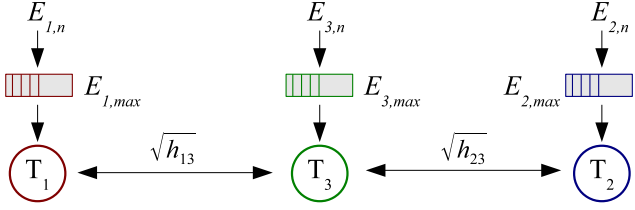}
\caption{The two-way relay channel with energy harvesting nodes (EH-TWRC).}
\label{fig_model}
\end{figure}

\section{System Model}
\label{sect_model}

We consider an additive white Gaussian noise (AWGN) two-way relay channel with two source nodes, $T_1$ and $T_2$, that convey independent messages to each other through a relay node $T_3$. The two source nodes cannot hear each other directly, hence all messages are sent through the relay. The channels to and from a source node are reciprocal\footnote{While this assumption is for the sake of simplicity, we note that the results of this paper directly extend to models without reciprocity.}, with power gains $h_{13}$ between nodes $T_1$ and $T_3$ and $h_{23}$ between nodes $T_2$ and $T_3$. We consider the delay limited scenario, where the relay forwards messages as soon as they are received, and thus has no data buffer. The channel model is shown in Figure~\ref{fig_model}.

\begin{figure}[t]
\centering
\includegraphics[width=\linewidth]{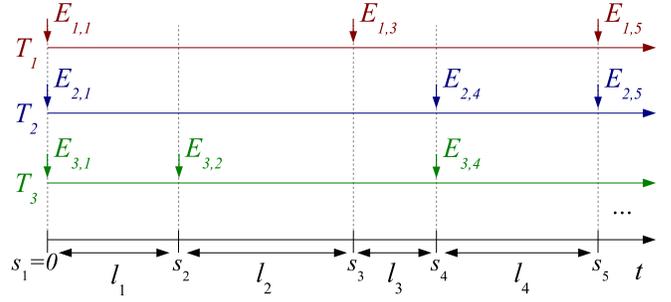}
\caption{The energy harvesting model for node $T_j$, $j=1,2,3$.}
\label{fig_energy_model}
\end{figure}

All nodes $T_1$, $T_2$ and $T_3$ are powered by energy harvesting. Node $T_j$, $j=1,2,3$, harvests $E_{j,n}\geq 0$ units of energy\footnote{Recent efforts extend the discrete energy arrivals to continuous ones for the single user channel and concludes similar insights albeit with a more involved analysis \cite{varan2014ita}.} at time $s_n$, and stores it in a battery of energy storage capacity $E_{j,max}$. Any energy in excess of the storage capacity of the battery is lost. The initial charge of the batteries are represented with $E_{j,1}$, with $s_1=0$ by definition. The time between the $n$th and $(n+1)$th energy arrivals, referred to as the $n$th epoch in the sequel, is denoted by $l_n=s_{n+1}-s_n$. We remark that the model does not require all nodes to harvest energy packets simultaneously; but rather indicates that epochs are constructed as the intervals between any two energy arrivals. A node that is not receiving any energy at the $n$th harvest is set to have $E_{j,n}=0$. The energy harvesting model is depicted in Figure~\ref{fig_energy_model}.

We consider a transmission session of $N$ epochs, with length $s_{N+1}$, for which the energy harvesting profile consists of $E_{j,n}$ and $s_n$ for $j=1,2,3$ and $n=1,\dots,N$. In epoch $n$, $n=1,\dots,N$, node $T_j$, $j=1,2,3$, allocates an average power $p_{j,n}$ for transmission, i.e., a total energy of $l_n p_{j,n}$ is consumed for transmitting. Since the energy available to each node is limited by the energy harvested and stored in the respective battery, the energy harvesting profile determines the feasibility of $p_{j,n}$ for each node. Specifically, the transmission powers satisfy
\begin{align}
\label{eqn_model_constraint}
p_{j,n} \leq \frac{B_{j,n}}{l_n}, \qquad j=1,2,3, ~~ n=1,\dots,N,
\end{align}
where $B_{j,n}$ is the energy available to node $j$ at the beginning of epoch $n$, which evolves as
\begin{align}
B_{j,n+1}=\min\{E_{j,max},B_{j,n} - l_n p_{j,n} + E_{j,n+1} \}.
\end{align}
In this work, similar to references \cite{yang2012optimal, tutuncuoglu2012optimum, ozel2011fading, ho2012optimal, devillers:imperfections, tutuncuoglu2014inefficient, tutuncuoglu2012ita, yang2012mac, yang2012broadcasting, ozel2012optimalbc, tutuncuoglu2012sum, antepli2011optimal, gunduz2011two, orhan2012optimal, orhan2012energy, orhan2013throughput, ahmed2012power, luo2012optimal2, huang2013throughput, ahmed2013joint}, the energy harvesting profile is known non-causally by all nodes, so that offline optimal policies and performance limits of the network can be found\footnote{We provide the online policy with causal energy arrival information in Section~\ref{sect_online}.}. The communication overhead for conveying energy arrival information and power allocation decisions is considered to be negligible compared to the amount of data transferred in each epoch.

We consider the problem of finding the power policy which maximizes the sum-throughput of the system under different relaying strategies such as decode-and-forward, compute-and-forward, compress-and-forward and amplify-and-forward. In the next subsection, we present the rate regions for these relaying strategies.

\subsection{Rate Regions with Average Power Constraints}
\label{sub_model_rates}

We focus on a two-phase communication scheme, consisting of a multiple access phase from nodes $T_1$ and $T_2$ to $T_3$ and a broadcast phase from $T_3$ to $T_1$ and $T_2$. This is referred to as multiple access broadcast (MABC) in \cite{kim2008performance, kim2011achievable}. Its three phase counterpart, time division broadcast (TDBC) \cite{kim2008performance, kim2011achievable}, can be shown to perform no better than MABC in the absence of a direct channel between $T_1$ and $T_2$, and is therefore omitted. For half-duplex nodes, the rates achievable with decode-and-forward, compress-and-forward, amplify-and-forward and compute-and-forward relaying schemes are derived in \cite{kim2008performance, kim2011achievable}. These works consider nodes that are constrained by their instantaneous transmit powers, and do not consider total consumed energy, which depends on the duration of multiple access and broadcast phases. Since our model is energy-constrained, we revise the results of these work by scaling transmit powers with phase duration, thereby replacing instantaneous transmit powers with average transmit powers $p_{j,n}$. We denote the set of rate pairs achievable with average transmit powers $p_1$, $p_2$ and $p_3$ and multiple access phase duration $\Delta$ as $\mathcal{R}_{HD}(p_1,p_2,p_3,\Delta)$ in the half-duplex case. The duration of the broadcast phase is $\bar\Delta=1-\Delta$. For full-duplex nodes, due to simultaneous multiple access and broadcast phases, there is no need for the time sharing factor $\Delta$; we use $\mathcal{R}_{FD}(p_1,p_2,p_3)$ to denote the achievable set of rate pairs. In this case, the full-duplex nodes can remove the self-interference term form the received signal, as in \cite{rankov2006achievable}. We use a subscript to denote the relaying strategy where necessary.

\begin{figure}[t]
\centering
\includegraphics[width=\linewidth]{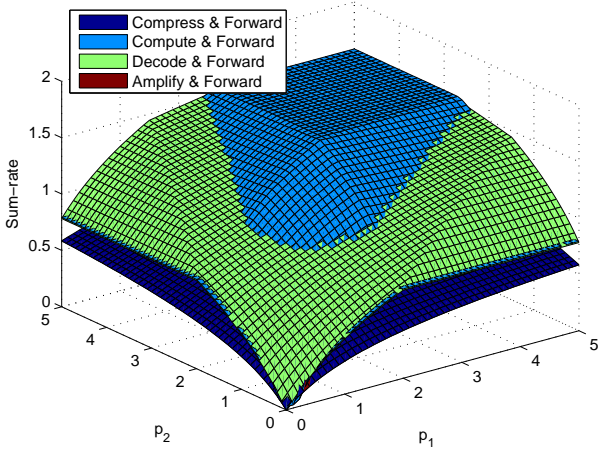}
\caption{Comparison of sum-rates for a symmetric full-duplex channel with $h_{13}=h_{23}=1$ at $p_{3}=2$. Amplify-and-forward rates remain just below compress-and-forward and thus are not visible.}
\label{fig_comp1_fd}
\end{figure}

{\bf Decode-and-Forward:} In this scheme, the relay decodes the messages of both source nodes in the multiple access phase, and transmits a function of the two messages in the broadcast phase. Nodes $T_1$ and $T_2$ use the broadcast message along with their own messages to find the ones intended for them. For half-duplex nodes, the rate region $\mathcal{R}_{DF-HD}(p_{1,n},p_{2,n},p_{3,n},\Delta_n)$ in epoch $n$ is defined by\footnote{The power gains, transmit powers and harvested energy values are normalized in order to obtain an effective noise variance of $1$ at each node. This is done by first scaling $h_{13}$ and $h_{23}$ to establish unit variance noise at nodes $T_1$ and $T_2$, and subsequently scaling the transmit power, available energy and battery capacity at nodes $T_1$ and $T_2$ to yield a unit variance noise at $T_3$.}
\begin{subequations}
\label{eqn_rates_hd_mabc}
\begin{align}
	& R_{1,n} \leq \min \left \{\Delta_n C \left (\frac{h_{13} p_{1,n}}{\Delta_n} \right),\bar{\Delta}_n C \left (\frac{h_{23} p_{3,n}}{\bar{\Delta}_n}
\right ) \right \}, \\
	& R_{2,n} \leq \min \left \{\Delta_n C \left (\frac{h_{23} p_{2,n}}{\Delta_n} \right),\bar{\Delta}_n C \left (\frac{h_{13} p_{3,n}}{\bar{\Delta}_n}
\right ) \right\}, \\
\label{eqn_rates_hd_mabc_sum}
	& R_{1,n}+R_{2,n} \leq \Delta_n C \left (\frac{h_{13} p_{1,n}+h_{23} p_{2,n}}{\Delta_n} \right ),
\end{align}
\end{subequations}
where $C(p)=\frac{1}{2}\log(1+p)$. With full-duplex radios, the two phases take place simultaneously, achieving instantaneous rates $(R_{1,n},R_{2,n}) \in \mathcal{R}_{DF-FD} (p_{1,n},p_{2,n},p_{3,n})$ which are found by substituting $\Delta_n=\bar{\Delta}_n=1$ in (\ref{eqn_rates_hd_mabc}).

{\bf Compress-and-Forward:} In this scheme, the relay transmits a compressed version of its received signal in the broadcast phase. The instantaneous rates $(R_{1,n},R_{2,n}) \in \mathcal{R}_{CF-HD}(p_{1,n},p_{2,n},p_{3,n},\Delta_n)$, $0 \leq \Delta_n \leq 1$, for the MABC half-duplex case satisfy
\begin{subequations}
\label{eqn_rates_hd_cf}
\begin{align}
\label{eqn_rates_hd_cf_1}
	R_{1,n} & \leq \Delta_n C \left ( \frac{(\sigma_{y}^{(1)})^2 h_{13} p_{1,n}/\Delta_n}{P_{\hat{y}}^{(1)}(P_{y}^{(1)})^2-(\sigma_{y}^{(1)})^2
(P_{y}^{(1)}-1)} \right ), \\	
\label{eqn_rates_hd_cf_2}
	R_{2,n} & \leq \Delta_n C \left ( \frac{(\sigma_{y}^{(1)})^2 h_{23} p_{2,n}/\Delta_n}{P_{\hat{y}}^{(1)}(P_{y}^{(1)})^2-(\sigma_{y}^{(1)})^2
(P_{y}^{(1)}-1)} \right ),
\end{align}
\end{subequations}
for some $P_{\hat{y}}^{(1)} \geq 0$ and $\sigma_{y}^{(1)} \geq 0$, where $P_y^{(1)}=h_{13} p_{1,n}/\Delta_n + h_{23}
p_{2,n}/\Delta_n +1$. The full-duplex rates $(R_{1,n},R_{2,n}) \in \mathcal{R}_{CF-FD}(p_{1,n},p_{2,n},p_{3,n})$ are \cite{rankov2006achievable}
\begin{align}
\label{eqn_rates_fd_cf}
	R_{1,n} \leq C \left ( \frac{h_{13} p_{1,n}}{1+\sigma_c^2} \right ), \qquad	
	R_{2,n} \leq C \left ( \frac{h_{23} p_{2,n}}{1+\sigma_c^2} \right ),
\end{align}
where $\sigma_c^2 = \max \{\sigma_{c1}^2,\sigma_{c2}^2\}$, and
\begin{subequations}
\label{eqn_rates_fd_cf_sigmas}
\begin{align}
	& \sigma_{c1}^2 = \frac{1+h_{23} p_{2,n}}{2^{2R_{3,n}}}, \quad	
	\sigma_{c2}^2 = \frac{1+h_{13} p_{1,n}}{2^{2R_{3,n}}},  \\
	& R_{3,n} = \min\{C(h_{13} p_{3,n}),C(h_{23} p_{3,n})\}.
\end{align}
\end{subequations}

{\bf Amplify-and-Forward:} In this scheme, the relay broadcasts a scaled version of its received signal. Since this is performed on a symbol-by-symbol basis, the time allocated for multiple access and broadcast phases are equal. The rate regions $\mathcal{R}_{AF}(p_{1,n},p_{2,n},p_{3,n})$ are found as
\begin{subequations}
\label{eqn_rates_fd_af}
\begin{align}
\label{eqn_rates_fd_af_1}
R_{1,n}\leq \Delta_n C\left(\frac{h_{13} h_{23} p_{1,n} p_{3,n}}{\Delta_n(h_{13} p_{1,n}+h_{23} (p_{2,n}+p_{3,n})+\Delta_n)}
\right), \\
\label{eqn_rates_fd_af_2}
R_{2,n}\leq \Delta_n C\left(\frac{h_{13} h_{23} p_{2,n} p_{3,n}}{\Delta_n(h_{23} p_{2,n}+h_{13} (p_{1,n}+p_{3,n})+\Delta_n)}
\right),
\end{align}
\end{subequations}
by substituting $\Delta_n=0.5$ for the half-duplex case and $\Delta_n=1$ for the full-duplex case.

{\bf Compute-and-Forward (Lattice Forwarding):} In this scheme, nested lattice codes are used at the source nodes, and the relay decodes and broadcasts a function of the two messages received from the sources. Each source then calculates the intended message using the side information of its own \cite{nam2010capacity}. The rate region $\mathcal{R}_{LF-HD}(p_{1,n},p_{2,n},p_{3,n},\Delta_n)$ achievable with this scheme for an MABC half-duplex relay consists of rates satisfying
\begin{subequations}
\label{eqn_rates_hd_lf}
\begin{align}
	R_{1,n}  \leq \min \bigg\{ &\frac{\Delta_n}{2} \log^+ \left( \frac{p_{1,n}}{p_{1,n}+p_{2,n}}+\frac{h_{13} p_{1,n}}{\Delta_n} \right), \nonumber \\ 
	& \bar{\Delta}_n C\left(\frac{h_{23} p_{3,n}}{\bar{\Delta}_n}\right) \bigg \}, \\	
	R_{2,n}  \leq \min \bigg \{ &\frac{\Delta_n}{2} \log^+ \left( \frac{p_{2,n}}{p_{1,n}+p_{2,n}}+\frac{h_{23} p_{2,n}}{\Delta_n} \right), \nonumber \\
	& \bar{\Delta}_n C\left(\frac{h_{13} p_{3,n}}{\bar{\Delta}_n}\right) \bigg \},
\end{align}
\end{subequations}
where $\bar{\Delta}_n=1-\Delta_n$ and $\log^+(x)=\max\{ \log x,0\}$. The full-duplex rate region $\mathcal{R}_{LF-FD}(p_{1,n},p_{2,n},p_{3,n})$ can be evaluated by substituting $\Delta_n=\bar{\Delta}_n=1$ in (\ref{eqn_rates_hd_lf}). In reference \cite{nam2010capacity}, it is shown that this strategy achieves within $\frac{1}{2}$ bits of TWRC capacity in each epoch.

\begin{figure}[t]
\centering
\includegraphics[width=\linewidth]{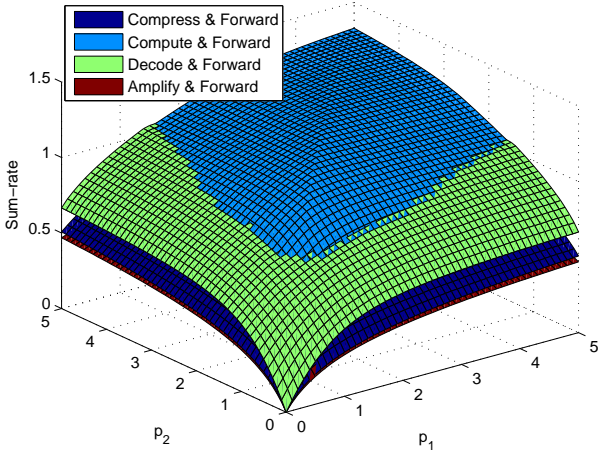}
\caption{Comparison of sum-rates for a symmetric half-duplex channel with $h_{13}=h_{23}=1$ at $p_{3}=2$. Amplify-and-forward rates remain just below compress-and-forward and thus are barely visible.}
\label{fig_comp1_hd}
\end{figure}

It can be observed that the compute-and-forward rates are not jointly concave in transmit powers $p_{j,n}$, $j=1,2,3$. This implies that time sharing between two sets of transmit powers $(p_{1,n},p_{2,n},p_{3,n})$ and $(\bar{p}_{1,n},\bar{p}_{2,n},\bar{p}_{3,n})$ with parameter $\lambda$, consuming average powers $\hat{p}_{j,n}=\lambda p_{j,n} + (1-\lambda)\bar{p}_{j,n}$, $j=1,2,3$, can yield rates $(R_{1,n},R_{2,n}) \notin \mathcal{R}_{LF-FD}(\hat{p}_{1,n},\hat{p}_{2,n},\hat{p}_{3,n})$. To include rates achievable as such, we concavify the rate region by extending $\mathcal{R}_{LF-FD}(p_{1,n},p_{2,n},p_{3,n})$ to include all time-sharing combinations with average power $(p_{1,n},p_{2,n},p_{3,n})$, i.e.,
\begin{align}
&\mathcal{R}_{LF-FD}^C (p_{1,n},p_{2,n},p_{3,n}) = \nonumber \\
&\qquad \Bigg\{  (R_{1,n},R_{2,n}) \Bigg | R_{k,n}= \sum_i \lambda_i R_{k,n,i}, \nonumber \\
&\qquad (R_{1,n,i},R_{2,n,i}) \in \mathcal{R}_{LF-FD}(p_{1,n,i},p_{2,n,i},p_{3,n,i}), \nonumber \\
&\qquad \sum_i \lambda_i=1, ~ \sum_i \lambda_i p_{j,n,i} \leq p_{j,n}, ~ \lambda_i \geq 0, \nonumber \\
&\qquad j=1,2,3, ~ k=1,2 \Bigg\},
\label{eqn_model_concavify}
\end{align}
which we refer to as the concavified rate region. This extends to the half-duplex relaying region $\mathcal{R}_{LF-HD}(p_{1,n},p_{2,n},p_{3,n})$ by time sharing among $\Delta_n$ as well. With a slight abuse of notation, we will denote the concavified regions with $\mathcal{R}_{LF-FD}(p_{1,n},p_{2,n},p_{3,n})$ and $\mathcal{R}_{LF-HD}(p_{1,n},p_{2,n},p_{3,n},\Delta_n)$ in the sequel. We note that all rates in the concavified region are achievable via time-sharing within an epoch, while the average powers within said epoch, and hence energy constraints, hold by definition. A formal proof of this concavification follows \cite[Lem.~1]{tutuncuoglu2012sum} closely. In the sequel, we use the concavified region, though we do not reiterate the required time-sharing for clarity of exposition. 

Since we are interested in maximizing sum-throughput, we compare the maximum achievable sum-rates for full- and half-duplex nodes employing the relaying strategies above in Figures~\ref{fig_comp1_fd} and \ref{fig_comp1_hd}, respectively. In these evaluations, a symmetric channel model normalized to yield $h_{13}=h_{23}=1$, and a fixed relay power of $p_{3}=2$ is considered. It can be observed that different schemes may outperform based on the instantaneous transmit power, and thus the selection of the correct relaying scheme is of importance in an energy harvesting setting where transmit powers are likely to change throughout the transmission.

\section{Hybrid Schemes}
\label{sect_hybrid}

In Section~\ref{sub_model_rates}, it is observed that depending on the transmit powers, either one of the relaying strategies may yield the best instantaneous sum-rate. Due to the intrinsic variability of harvested energy, transmit powers may change significantly throughout the transmission period based on the energy availability of nodes. Consequently, a dynamic relay that chooses its relaying strategy based on instantaneous transmit powers of the nodes can potentially improve system throughput.

Another benefit of switching between relaying strategies is achieving time-sharing rates across strategies, e.g., switching between decode-and-forward and compute-and-forward strategies within an epoch, which can outperform both individual strategies with the same average power. An example of the benefits of time-sharing in a two-way relay channel is reference \cite{liu2012optimizing}, where time-sharing between different operation modes is considered. In \cite{liu2012optimizing}, a fixed relaying strategy is employed with different nodes transmitting at a time; while here we allow time-sharing between different relaying strategies.

The rates achievable with a hybrid strategy switching between the four relaying schemes in Figures~\ref{fig_comp1_fd} and \ref{fig_comp1_hd} consist of the convex hull of the union of rate pairs achievable by the individual schemes. The rate region for the hybrid scheme is expressed as
\begin{align}
\nonumber
&\mathcal{R}_{HYB}(p_{1},p_{2},p_{3}) = \Bigg \{ (R_1,R_2) \Bigg |
R_k=\sum_i \lambda_i R_{k,i}, \\
&\quad~~ \sum_i \lambda_i = 1,~ \sum_i \lambda_i p_{j,i} \leq p_{j}, ~ \lambda_i \geq 0, \nonumber \\
\label{eqn_sumrate_hybrid}
&\quad~~ (R_{1,i},R_{2,i}) \in \mathcal{R}_{DF} \cup \mathcal{R}_{LF} \cup \mathcal{R}_{CF} \nonumber \\
&\quad~~ \cup \mathcal{R}_{AF}(p_{1,i},p_{2,i},p_{3,i}), ~ j=1,2,3, ~k=1,2 \Bigg \},
\end{align}
where $\mathcal{R}_{DF}$, $\mathcal{R}_{LF}$, $\mathcal{R}_{CF}$ and $\mathcal{R}_{AF}$ are the rate regions given in Section~\ref{sub_model_rates} with decode-and-forward, compute-and-forward, compress-and-forward and amplify-and-forward, respectively.

\begin{figure}[t]
\centering
\includegraphics[width=0.95\linewidth]{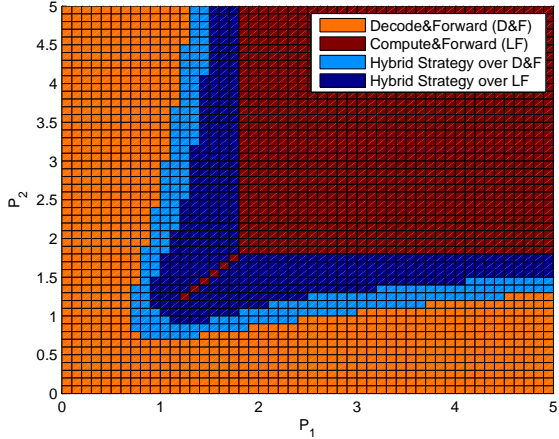}
\caption{Chosen relaying strategy for a symmetric half-duplex channel with $h_{13}=h_{23}=1$ at $p_{3}=2$. The labels ``over D\&F'' and ``over LF'' denote which of the two strategies is better by itself in that region.}
\label{fig_sims_hybrid}
\end{figure}

For the purpose of demonstration, we present the chosen relaying scheme that maximizes the instantaneous sum-rate for a half-duplex channel with fixed relay transmit power, $p_3=2$, in Figure~\ref{fig_sims_hybrid}. It can be observed that while decode-and-forward or compute-and-forward alone are chosen at the extremes, a time-sharing of the two strategies is favored in between. In this figure, the regions where the hybrid scheme uses time-sharing are shown in two shades of blue. We note that for these channel parameters, the remaining relaying schemes under-perform these two for any choice of transmit powers.

With these observations, we conclude that policies with hybrid relaying strategies can instantaneously surpass the sum-rates resulting from individual relaying schemes for a considerable set of power vectors. Furthermore, time-sharing between relaying strategies may strictly outperform the best relaying strategy alone. Numerical results on the performance of optimal hybrid schemes in comparison with individual schemes are presented in Section~\ref{sect_numerical}.

\section{Problem Definition and Properties of the Optimal Solution}
\label{sect_properties}

We consider the problem of sum-throughput maximization for a session of $N$ epochs. Since achievable rates are either jointly concave in transmit powers or can be concavified by the use of time sharing as in (\ref{eqn_model_concavify}), it follows that the optimal transmit powers remain constant within each epoch, as noted in \cite[Lemma 2]{yang2012optimal}. The power policy of the network consists of the power vectors $(\bb{p}_1,\bb{p}_2,\bb{p}_3)$, where $\bb{p}_j = (p_{j,1},p_{j,2},\dots,p_{j,N})$, $j=1,2,3$, and in the case of half-duplex relaying, the time sharing parameters $\Delta_n$, $n=1,\dots,N$. For the set of feasible power policies, we first present the following proposition, which is the multi-user extension of \cite[Lemma 2]{tutuncuoglu2012optimum}:

\begin{proposition}
\label{lem_overflow}
	There exists optimal average transmit powers $(\bb{p}_1^*,\bb{p}_2^*,\bb{p}_3^*)$ that do not yield a battery overflow at any of the nodes throughout the communication session.
\end{proposition}
\begin{Proof}
Let $(\bb{p}_1,\bb{p}_2,\bb{p}_3)$ be a vector of transmit powers yielding battery overflows, i.e., 
\begin{align}
\sum_{i=1}^{n} E_{j,i} - \sum_{i=1}^{n-1} l_i p_{j,i} - E_{j,max} = E_{j,n}^{ovf} > 0
\end{align}
for some $j$ and $n$. For each battery overflow of amount $E_{j,n}^{ovf}$ at node $T_j$ at the end of epoch $n$, let $\bar{p}_{j,n}=p_{j,n}+\tfrac{E_{j,n}^{ovf}}{l_n}$. For the remaining powers, let $\bar{p}_{j,n}=p_{j,n}$. The power policy defined by $(\bb{\bar{p}}_1,\bb{\bar{p}}_2,\bb{\bar{p}}_3)$ does not overflow the battery at any time, and satisfies $\bar{p}_{j,n} \geq p_{j,n}$ for all $j$ and $n$. Note that nodes consuming powers $\bar{p}_j$ can achieve any rate pair that is achievable with less power, i.e.,
\begin{subequations}
\label{eqn_prop_nondiminishing}
\begin{align}
  & p_{j} \leq \bar{p}_{j}, j=1,2,3 \nonumber \\
  &~~ \Rightarrow \mathcal{R}_{FD}(p_{1},p_{2},p_{3}) \subset \mathcal{R}_{FD}(\bar{p}_{1},\bar{p}_{2},\bar{p}_{3}), \\
    &~~ \Rightarrow \mathcal{R}_{HD}(p_{1},p_{2},p_{3},\Delta) \subset \mathcal{R}_{HD}(\bar{p}_{1},\bar{p}_{2},\bar{p}_{3},\Delta),
\end{align}
\end{subequations}
for full-duplex and half-duplex nodes with $0 \leq \Delta \leq 1$, respectively. Therefore, the sum-rate obtained by $(\bb{\bar{p}}_1,\bb{\bar{p}}_2,\bb{\bar{p}}_3)$ at any epoch $n$ is no less than that of $(\bb{p}_1,\bb{p}_2,\bb{p}_3)$. Hence, for any policy with battery overflows, we can find a policy performing at least as good without overflows.
\end{Proof}

We remark that even though (\ref{eqn_prop_nondiminishing}) does not hold immediately, e.g., for the amplify-and-forward rates in (\ref{eqn_rates_fd_af}), it holds by definition for the concavified rates in (\ref{eqn_model_concavify}). By choosing $\lambda_1=1$ and $p_{j,n,1}<p_{j,n}$ in (\ref{eqn_model_concavify}), a portion of the allocated power $p_{j,n}$ can equivalently be discarded at the node. Consequently, Proposition~\ref{lem_overflow} applies to all concavified relaying schemes presented in Section~\ref{sub_model_rates}.

As a consequence of Proposition~\ref{lem_overflow}, we will restrict the feasible set of policies to those that do not overflow the battery without loss of generality. In epoch $n$, the nodes choose transmit powers $(p_{1,n},p_{2,n},p_{3,n})$, a time sharing parameter $\Delta_n$, and a rate pair $(R_{1,n},R_{2,n}) \in \mathcal{R}_{HD}(p_{1,n},p_{2,n},p_{3,n},\Delta_n)$ in the case of half-duplex radios. The objective is to maximize the sum-throughput of the TWRC within $N$ epochs, where the transmit powers are constrained by harvested energy and the rates are constrained by the rate region. We express the EH-TWRC sum-throughput maximization problem
\begin{subequations}
\label{eqn_model_prob_hd}
\begin{align}
\label{eqn_model_hd_objective}
	& \max_{\bb{R}_1,\bb{R}_2,\bb{p}_1,\bb{p}_2,\bb{p}_3,\bb{\Delta}}  ~~ \sum_{i=1}^{N} l_i (R_{1,i} + R_{2,i}) \\
\label{eqn_model_hd_c1}	
	&~~ \mbox{s.t.} ~~(R_{1,n},R_{2,n}) \in \mathcal{R}_{HD}(p_{1,n},p_{2,n},p_{3,n},\Delta_n),  \\
\label{eqn_model_hd_causality}
	&\qquad~~ \sum_{i=1}^n l_i p_{j,i} - \sum_{i=1}^n E_{j,i} \leq 0, \\
\label{eqn_model_hd_capacity}
	&\qquad~~ \sum_{i=1}^{n} E_{j,i} - \sum_{i=1}^{n-1} l_i p_{j,i} \leq E_{j,max}, \\
	&\qquad~~   0 \leq \Delta_n \leq 1, ~ j=1,2,3, ~ n=1,2,\dots,N,
\end{align}
\end{subequations}
for half-duplex nodes, where $\bb{p}_j = (p_{j,1},p_{j,2},\dots,p_{j,N})$, $j=1,2,3$, $\bb{R}_k = (R_{k,1},R_{k,2},\dots,R_{k,N})$, $k=1,2$, and $\bb{\Delta} = (\Delta_1,\Delta_2,\dots,\Delta_N)$. Here, (\ref{eqn_model_hd_capacity}) is due to Proposition~\ref{lem_overflow}, and (\ref{eqn_model_hd_causality}) is equivalent to (\ref{eqn_model_constraint}) given (\ref{eqn_model_hd_capacity}). While the rates are a function of the powers of the nodes and the time sharing parameters $\Delta_n, n=1,2,\dots,N$, this dependency is now deferred to (\ref{eqn_model_hd_c1}), which is the constraint that ensures the rates are selected from the achievable region dictated by the power and time sharing parameters. The energy causality constraints given in (\ref{eqn_model_hd_causality}) ensure that the energy consumed by a node is not greater than the energy harvested up to that epoch. The no-overflow constraints given in (\ref{eqn_model_hd_capacity}) ensure that the battery capacity is not exceeded. Any power policy $(\bb{p}_1,\bb{p}_2,\bb{p}_3)$ satisfying both (\ref{eqn_model_hd_causality}) and (\ref{eqn_model_hd_capacity}) for all $j$ and $n$ is considered a feasible power power policy. The problem for full-duplex nodes is attained by replacing (\ref{eqn_model_hd_c1}) with $(R_{1,n},R_{2,n}) \in \mathcal{R}_{FD} (p_{1,n},p_{2,n},p_{3,n})$ and omitting the time-sharing variables $\Delta_n$, $n=1,\dots,N$.

We next show that (\ref{eqn_model_prob_hd}) can be decomposed by separating the maximization over $p_{1,n}$, $p_{2,n}$, and $p_{3,n}$, $n=1,\dots,N$, and the maximization over $R_{1,n}$, $R_{2,n}$, $\Delta_n$, $n=1,\dots,N$, as
\begin{subequations}
\label{eqn_model_prob_hd2}
\begin{align}
\label{eqn_model_hd2_objective}
	& \max_{\bb{p}_1,\bb{p}_2,\bb{p}_3} ~~ \max_{\bb{R}_1,\bb{R}_2,\bb{\Delta}} ~~ \sum_{i=1}^{N} l_i (R_{1,i} + R_{2,i}) \\
\label{eqn_model_hd2_c1}	
	&~~ \mbox{s.t.}  ~~(R_{1,n},R_{2,n}) \in \mathcal{R}_{HD}(p_{1,n},p_{2,n},p_{3,n},\Delta_n), \\
\label{eqn_model_hd2_causality}
	&\qquad~~ \sum_{i=1}^n l_i p_{j,i} - \sum_{i=1}^n E_{j,i} \leq 0, \\
\label{eqn_model_hd2_capacity}
	&\qquad~~ \sum_{i=1}^{n} E_{j,i} - \sum_{i=1}^{n-1} l_i p_{j,i} \leq E_{j,max}, \\
    &\qquad~~ 0 \leq \Delta_n \leq 1, ~	 j=1,2,3, ~ n=1,2,\dots,N.
\end{align}
\end{subequations}
Note that only the constraints in (\ref{eqn_model_hd2_c1}) pertain to the parameters of the second maximization. Next, we observe that the constraints in (\ref{eqn_model_hd2_c1}) are separable in $n$, and the objective is a linear function of $R_{1,n}$ and $R_{2,n}$. Hence, the second maximization can be carried out separately for each $n$, i.e., in an epoch-by-epoch fashion, yielding the separated problem
\begin{subequations}
\label{eqn_model_prob_norate}
\begin{align}
\label{eqn_model_prob_norate_objective}
	\max_{\bb{p}_1,\bb{p}_2,\bb{p}_3} &~~ \sum_{i=1}^{N} l_i R_s(p_{1,i},p_{2,i},p_{3,i}) \\
\label{eqn_model_prob_norate_causality}	
	\mbox{s.t.} ~ &~~\sum_{i=1}^n l_i p_{j,i} - \sum_{i=1}^n E_{j,i} \leq 0, &\\
\label{eqn_model_prob_norate_capacity}
	&~~ \sum_{i=1}^{n} E_{j,i} - \sum_{i=1}^{n-1} l_i p_{j,i} \leq E_{j,max}, \\
	&~~  j=1,2,3, ~ n=1,2,\dots,N,
\end{align}
\end{subequations}
where $R_s(p_{1,i},p_{2,i},p_{3,i})$ is the solution to
\begin{subequations}
\label{eqn_model_prob_hd_inner}
\begin{align}
\label{eqn_model_hd_inner_objective}
& \max_{R_{1,i},R_{2,i},\Delta_i} ~~ R_{1,i} + R_{2,i} &\\
\label{eqn_model_hd_inner_c1}	
	&\qquad~ \mbox{s.t.} ~~ (R_{1,i},R_{2,i}) \in \mathcal{R}_{HD}(p_{1,i},p_{2,i},p_{3,i},\Delta_i), \\
	&\qquad \qquad~ 0 \leq \Delta_i \leq 1,
\end{align}
\end{subequations}
within a single epoch $i$ with fixed powers $(p_{1,i},p_{2,i},p_{3,i})$. This implies that the optimal transmit rates within each epoch are the sum-rate maximizing rates for the given transmit powers within that epoch. Thus, we refer to the function $R_s(p_{1,i},p_{2,i},p_{3,i})$ as the {\em maximum epoch sum-rate}. For full-duplex nodes, the maximum epoch sum-rate is found by solving
\begin{subequations}
\label{eqn_model_prob_fd_inner}
\begin{align}
\label{eqn_model_fd_inner_objective}
	\max_{R_{1,i},R_{2,i}} &~~ R_{1,i} + R_{2,i} &\\
\label{eqn_model_fd_inner_c1}	
	\mbox{s.t.} & ~~(R_{1,i},R_{2,i}) \in \mathcal{R}_{FD}(p_{1,i},p_{2,i},p_{3,i})
\end{align}
\end{subequations}
instead, and the power policy optimization is identical to (\ref{eqn_model_prob_norate}). We next show a property of policies that solve the problem in (\ref{eqn_model_prob_hd}).

\begin{lemma}
\label{lem_depleted_batteries} There exists an optimal policy which depletes the batteries of all nodes at the end of transmission.
\end{lemma}
\begin{Proof}
Let $(\bb{R}_1,\bb{R}_2,\bb{p}_1,\bb{p}_2,\bb{p}_3)$ be a transmission policy which leaves energy $\mathcal{E}_j$ in the battery of node $j$ at the end of transmission. Consider the transmission policy $(\bar{\bb{R}}_1,\bar{\bb{R}}_2,\bar{\bb{p}}_1,\bar{\bb{p}}_2,\bar{\bb{p}}_3)$ which has $\bar{p}_{j,N}=p_{j,N}+\mathcal{E}_j/{l_{N}}$, and equals the original policy elsewhere. Hence, this policy expends the remaining energy in the battery of $T_j$ in the last epoch, depleting the batteries. We have $\bar{R}_{k,n} = R_{k,n}$ for $n=1,2,\dots,N-1$ and $\bar{R}_{k,N} \geq R_{k,N}$, for $k=1,2$, due to (\ref{eqn_prop_nondiminishing}). Therefore, the sum-throughput of the new policy cannot be lower than that of the original policy.
\end{Proof}

\section{Identifying the Optimal Policy}
\label{sect_identify}

Now that we have formulated the problem and identified some necessary properties of the optimal policy, we next find the optimal power policy for the EH-TWRC. We establish this using a generalization of the directional water-filling algorithm in \cite{ozel2011fading}, which gives the optimal policy for a single transmitter fading channel. In this section, we show the optimality of the generalized directional water-filling algorithm and verify its convergence.

\subsection{Solution of the EH-TWRC Sum-Throughput Maximization Problem}
\label{sub_solution}

To find the optimal policy, we first find the maximum epoch sum-rate by solving (\ref{eqn_model_prob_hd_inner}) and (\ref{eqn_model_prob_fd_inner}) for half-duplex and full-duplex nodes, respectively. The following property of $R_s(p_{1,i},p_{2,i},p_{3,i})$ can be immediately observed for any relaying scheme.

\begin{lemma}
\label{lem_sumrate_convex} The maximum epoch sum-rate $R_s(p_{1,i},p_{2,i},p_{3,i})$ is jointly concave in transmit powers $p_{1,i}$,
$p_{2,i}$, and $p_{3,i}$.
\end{lemma}
\begin{Proof}
Proof follows from the concavity of objectives (\ref{eqn_model_hd_inner_objective}) and (\ref{eqn_model_fd_inner_objective}), and the convexity of constraint sets (\ref{eqn_model_hd_inner_c1}) and (\ref{eqn_model_fd_inner_c1}). Let $(R_1,R_2)$ and $(\tilde{R}_1,\tilde{R}_2)$ denote two feasible rate pairs, and $\Delta$ and $\tilde\Delta$ their time-sharing parameters for transmit powers $(p_{1},p_{2},p_{3})$ and $(\tilde{p}_{1},\tilde{p}_{2},\tilde{p}_{3})$, respectively. Let $\bar{R}_k=\alpha R_k+(1-\alpha)\tilde{R}_k$, $k=1,2$, $\bar{p}_j=\alpha p_{j}+(1-\alpha)\tilde{p}_j$, and $\bar{\Delta}=\alpha \Delta+(1-\alpha)\tilde{\Delta}$, $j=1,2,3$ denote the convex combination of the policies with parameter $0 \leq \alpha \leq 1 $. Then, for all relaying schemes, $(\bar{R}_1,\bar{R}_2) \in \mathcal{R}_{FD}(\bar{p}_1,\bar{p}_2,\bar{p}_3)$ or $(\bar{R}_1,\bar{R}_2) \in \mathcal{R}_{HD}(\bar{p}_1,\bar{p}_2,\bar{p}_3,\bar{\Delta})$ follows either from the definition of the rate region, or from (\ref{eqn_model_concavify}).
\end{Proof}

As a consequence of Lemma~\ref{lem_sumrate_convex}, (\ref{eqn_model_prob_norate}) is a convex program. We next provide the \emph{iterative generalized directional water-filling algorithm} to compute the optimal power policy. Consider the power allocation problem in (\ref{eqn_model_prob_norate}) for an arbitrary relaying scheme with the maximum epoch sum-rate $R_s(p_{1},p_{2},p_{3})$. Here, the constraints in (\ref{eqn_model_prob_norate_causality}) and (\ref{eqn_model_prob_norate_capacity}) are separable among $j=1,2,3$. Hence, a block coordinate descent algorithm, i.e., alternating maximization, can be employed \cite{bertsekas1999nonlinear}. In each iteration, the power allocation problem for node $T_j$, $j=1,2,3$, given by
\begin{subequations}
\label{eqn_kkt}
\begin{align}
\label{eqn_iterative_objective}
	\max_{\bb{p}_j \geq 0} &~~ \sum_{n=1}^N l_n R_s(p_{1,n},p_{2,n},p_{3,n}) \\
\label{eqn_iterative_c1}	
	\mbox{s.t.} &~~\sum_{i=1}^n l_i p_{j,i} - \sum_{i=1}^n E_{j,i} \leq 0, \\
	&~~ \sum_{i=1}^{n} E_{j,i} - \sum_{i=1}^{n-1} l_i p_{j,i} \leq E_{j,max}, \\
	&~~ n=1,2,\dots,N,
\end{align}
\end{subequations}
is solved while keeping the remaining power levels $\bb{p}_k$, $k \neq j$, constant. This is a convex single user problem, and the solution satisfies the KKT stationarity conditions and complementary slackness conditions \cite{bertsekas1999nonlinear}
\begin{subequations}
\label{eqn_rates_gidwf_iter}
\begin{align}
\label{eqn_iterative_stationarity}
l_n \frac{\partial R_s(p_{1,n},p_{2,n},p_{3,n})}{\partial p_{j,n}} - l_n \sum_{i=n}^N (\lambda_i-\beta_i) + \gamma_n &=0, \\
\label{eqn_iterative_cs1}
\lambda_n \left( \sum_{i=1}^n l_i p_{j,i} - \sum_{i=1}^n E_{j,i} \right) &= 0, \\
\label{eqn_iterative_cs2}
\beta_n \left( \sum_{i=1}^{n} E_{j,i} \!-\! \sum_{i=1}^{n-1} l_i p_{j,i} \!-\! E_{j,max} \right) \!=\! 0, ~ \gamma_n p_{j,n} \!&=\!0,
\end{align}
\end{subequations}
for all $n=1,\dots,N$ where $\lambda_n \geq 0$, $\beta_n \geq 0$ and $\gamma_n \geq 0$ are the Lagrange multipliers for energy causality, battery capacity and transmit power non-negativity constraints, respectively. Hence, the optimal transmit power policy for $T_j$, i.e., $\bb{p}_j$, is the solution to
\begin{equation}
\label{eqn_water-level2}
\frac{\partial R_s(p_{1,n},p_{2,n},p_{3,n})}{\partial p_{j,n}} = \sum_{i=n}^N (\lambda_i-\beta_i) - \frac{\gamma_n}{l_n}
\end{equation}
for all $n=1,...,N$ which follows from (\ref{eqn_iterative_stationarity}). Note that due to (\ref{eqn_iterative_cs1}) and (\ref{eqn_iterative_cs2}), the Lagrange multipliers are nonzero only when the respective constraints are met with equality. 

\begin{figure}
\centering
\includegraphics[width=\linewidth]{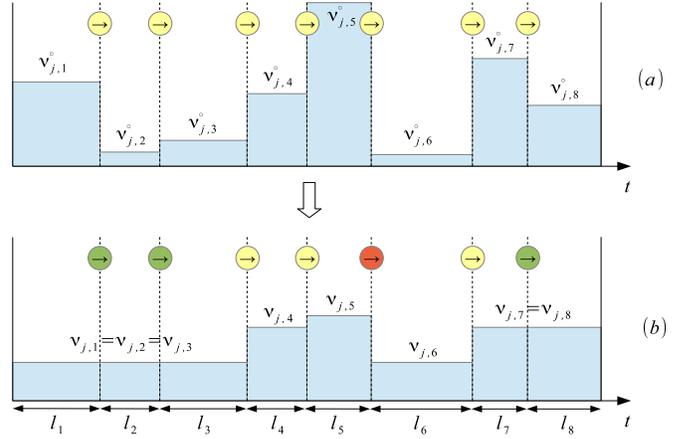}
\caption{Depiction of generalized directional water-filling for $T_j$ with $N=8$ epochs. Note that the battery of the node is full at the end of the 5th epoch, preventing further energy flow into the 6th epoch.}
\label{fig_wf}
\end{figure}

We argue that the solution to (\ref{eqn_water-level2}) can be interpreted as a generalization of the directional water-filling algorithm \cite{ozel2011fading} similar to the case in \cite{tutuncuoglu2012sum}. In \cite{ozel2011fading}, optimal transmit powers are found by treating the available energy in each epoch as water, and letting water levels equalize by flowing in the forward direction only. The associated algorithm is termed directional water-filling. Here, we instead define the {\em generalized water levels} for $T_j$ as 
\begin{equation}
\label{eqn_water-level}
\nu_{j,n}(p_{j,n})=\left( \frac{\partial R_s(p_{1,n},p_{2,n},p_{3,n})}{\partial p_{j,n}} \right) ^{-1}.
\end{equation}
The following properties of $\nu_{j,n}$ are readily observed for the optimal policy: (a) while $p_{j,n} > 0$, the water levels remain constant among epochs unless the battery is empty or full, increasing only when the battery is empty, and decreasing only when the battery is full, and (b) if a positive solution to 
\begin{equation}
\label{eqn_property_a}
\frac{\partial R_s(p_{1,n},p_{2,n},p_{3,n})}{\partial p_{j,n}} = \sum_{i=n}^N (\lambda_i-\beta_i)
\end{equation}
does not exist, then $p_{j,n}=0$ and $\gamma_n \geq 0$. These properties imply that the optimal policy can be found by performing directional water-filling using the generalized water levels in (\ref{eqn_water-level}), and calculating the corresponding transmit powers $p_{j,n}$. Water flow is only in the forward direction and the corresponding energy flow is bounded by $E_{j,max}$ for node $T_j$. Hence, the flow between two neighboring epochs stops when water levels in (\ref{eqn_water-level}) are equalized or when the total energy flow reaches $E_{j,max}$. The initial water levels are found by substituting the initial transmit powers $p_{j,n}^\circ=E_{j,n}/l_n$ in (\ref{eqn_water-level}). This algorithm yields transmit powers that satisfy the two properties above by construction. An example of generalized directional water-filling is depicted in Figure~\ref{fig_wf}.


The {\em iterative} generalized directional water-filling (IGDWF) algorithm employs generalized directional water-filling sequentially for each user until all power levels $\bb{p}_j$, $j=1,2,3$, converge, i.e., alternating maximization. Although optimization is carried on separately for a single user at each iteration, the transmit powers of all users interact through the generalized water levels in (\ref{eqn_water-level}). Starting from the initial values $p_{j,n}^{(0)}=E_{j,n}/l_n$, the $k$th iteration of the algorithm, optimizing $\bb{p}_j^{(k)}$ for $j=(k~\mbox{mod}~3)+1$, is given in Algorithm~\ref{alg_1}.

\begin{remark}
At each iteration of the IGDWF algorithm, the water flow out of each of the $N$
epochs can be found using a binary search. This requires updating at most $N$ water levels following each epoch. Hence, the computational complexity of each iteration is $O(N^2)$, i.e., quadratic in the number of epochs.
\end{remark}

\begin{algorithm}[t]
  
 1) Let $j=(k~\mbox{mod}~3)+1$, $p_{j,n}^{(k)} = p_{j,n}^{(0)}$, $p_{\ell,n}^{(k)} = p_{\ell,n}^{(k-1)}$ for $\ell \neq j$, $\delta_n = E_{j,n}$, $n=1,...,N$.
 
 2) {\bf for} $n=2,...,N$, {\bf do}
 
 ~~~~ Find the set $\mathcal{E} = \big\{ E_\Delta \geq 0 \big | \nu_{j,n-1}(p_{j,n-1}^{(k)}-\frac{E_\Delta}{l_{n-1}})$ 
 
 ~~~~ $=\nu_{j,n}(p_{j,n}^{(k)}+\frac{E_\Delta}{l_{n}}),~\delta_n+E_\Delta \leq E_{j,max} \big\}$,
 
 ~~~~ {\bf if} $\mathcal{E} = \emptyset$ and $\nu_{j,n-1}(p_{j,n-1}^{(k)}) > \nu_{j,n}(p_{j,n}^{(k)})$, {\bf then}  
 
 ~~~~~~~ assign $\mathcal{E} = \{ E_{j,max}-\delta_n \}$,
 
 ~~~~ Find $E_\Delta \in \mathcal{E}$ and assign $p_{j,n-1}^{(k)} = p_{j,n-1}^{(k)}-\frac{E_\Delta}{l_{n-1}}$, 
 
 ~~~~ $p_{j,n}^{(k)} = p_{j,n}^{(k)}+\frac{E_\Delta}{l_{n}}$, $\delta_n = \delta_n+E_\Delta$ 
 
 ~~~~ such that $\|\bb{p}_j^{(k)}-\bb{p}_j^{(k-1)}\|$ is minimized.
  
 ~~ {\bf end for}
 
 3) Repeat 2 until $\nu_{j,n-1}(p_{j,n-1}^{(k)}) \leq \nu_{j,n}(p_{j,n}^{(k)})$ or $\delta_n = E_{j,max}$ for all $n$.
  
 \caption{Iteration $k$ of Iterative Generalized Directional Water-filling} 
 \label{alg_1}
 
\end{algorithm}

\subsection{Convergence of the IGDWF Algorithm}

For the alternating maximization in Section~\ref{sub_solution} to converge to an optimal policy, it is sufficient that the feasible set is the intersection of convex constraints that are separable among $j=1,2,3$, and the continuously differentiable objective yields a unique maximum in each iteration \cite[Prop. 2.7.1]{bertsekas1999nonlinear}. In this case, the objective in (\ref{eqn_model_prob_norate_objective}) is concave and continuously differentiable for all relaying strategies, with compute-and-forward satisfying this condition after the concavification in (\ref{eqn_model_concavify}). The feasible set (\ref{eqn_model_prob_norate_causality})-(\ref{eqn_model_prob_norate_capacity}) is separable among $j=1,2,3$ as well. However, the objective does not necessarily yield a unique maximum at each iteration since it is not strictly concave in transmit powers. To overcome this, we introduce the unconstrained variables $\bb{s}_j=(s_{j,1},\dots,s_{j,N})$ for $j=1,2,3$, and modify the objective in (\ref{eqn_iterative_objective}) as
\begin{align}
\label{eqn_new_objective}
&f(\bb{p}_1,\bb{p}_2,\bb{p}_3,\bb{s}_1,\bb{s}_2,\bb{s}_3)=\sum_{n=1}^N l_n R_s(p_{1,n},p_{2,n},p_{3,n}) \nonumber \\
&~~ -\epsilon_1 \|\bb{p}_1-\bb{s}_1\|^2-\epsilon_2 \|\bb{p}_2-\bb{s}_2\|^2-\epsilon_3 \|\bb{p}_3-\bb{s}_3\|^2,
\end{align}
where $\epsilon_j > 0$, $j=1,2,3$ are arbitrarily small parameters. The objective in (\ref{eqn_new_objective}) is maximized by a unique $\bb{p}_j$ in each iteration with $j=1,2,$ or $3$. The iterations optimizing $\bb{s}_j$ trivially yield the unique solution $\bb{s}_j=\bb{p}_j$. Therefore, the problem now satisfies the convergence property for alternating maximization, and converges to the global maximum of (\ref{eqn_model_prob_norate}) \cite[Ex. 2.7.2]{bertsekas1999nonlinear}.

Note that through (\ref{eqn_new_objective}) and the arbitrarily small $\epsilon_j$, we essentially introduce {\em resistance} to the iterative algorithm. That is, if the original objective in (\ref{eqn_model_prob_norate_objective}) yields multiple solutions for some $j$, the objective in (\ref{eqn_new_objective}) has a unique solution that is closest to the previous value of $\bb{p}_j$. Consequently, if there exists more than one optimal solution to (\ref{eqn_kkt}) at one of the iterations for some $j$, the power policy $\bb{p}_j$ that is closest to the previous one is chosen. This is ensured by choosing the flow amount $E_\Delta$ which minimizes $\| \bb{p}_j^{(k)}-\bb{p}_j^{(k-1)}\|$ in Step~2 of Algorithm~\ref{alg_1}.


\section{Online Power Policy with Dynamic Programming}
\label{sect_online}

The power allocation policy we have considered so far is an offline policy, in the sense that the energy harvest amounts and times are known to all nodes in advance. Although the offline approach is useful for predictable energy harvesting scenarios \cite{gorlatova2011networking} and as a benchmark, it is also meaningful to develop policies that only rely on past and current energy states, i.e., causal information only. We refer to such transmission policies as {\it online} policies. Recent efforts that consider online algorithms for energy harvesting nodes in various channel models include \cite{ho2012optimal, ozel2011fading, blasco2013low, blasco2012learning, khuzani2013optimal, ahmed2012power, li2011relay}. Building upon the previous work, in this section, we find the optimal online policy for power allocation in the two-way relay channel.

The epoch length $l_n$ indicates that no energy will be harvested for a duration of $l_n$ after the $n$th energy arrival. Therefore, in the online problem, the epoch lengths are not known by the nodes causally. Instead, we divide the transmission period into time slots of length $\tau$, and recalculate transmit powers at the beginning of each time slot. We assume that each energy harvest takes place at the beginning of some time slot. Note that with smaller $\tau$, this model gets arbitrarily close to the general model in Section~\ref{sect_model}. We assume that harvests $E_{j,n}$ in time slot $n$ are independent and identically distributed. In time slot $n$, nodes $T_1$, $T_2$ and $T_3$ have access to previous energy harvests $E_{j,i}$, $j=1,2,3,$ and $i=1,...,n$. The nodes decide on transmit powers $p_{j,n}$, $j=1,2,3,$ through {\em actions} 
\begin{align}
\label{eqn_online_action1}
	p_{j,n}=\phi_{j,n}(\{E_{k,i};k=1,2,3,i=1,...,n\}),
\end{align}
where $\{E_{k,i};k=1,2,3,i=1,...,n\}$ denotes all energy arrivals prior to, and including, time slot $n$. Each time slot with transmit powers $\{p_{j,n}\}$ contribute to the additive objective through the sum-rate function $R_S(p_{1,n},p_{2,n},p_{3,n})$ in (\ref{eqn_model_prob_hd_inner}) and (\ref{eqn_model_prob_fd_inner}) for full-duplex and half-duplex modes, respectively. We consider the problem of finding the optimal set of actions for this setting, which can be formulated as the following dynamic program \cite{bertsekas1995dynamic}:
\begin{align}
\label{eqn_online_bellman1}
	V(\{E_{j,i}\}_{i=1}^n) =& \underset{\phi_{1,n},\phi_{2,n},\phi_{3,n}}{\max}  R_s(\phi_{j,n}(\{E_{k,i}\})) \nonumber \\
	&\qquad + \mathbb{E} \left[ \sum_{i=n+1}^N V(\{E_{j,m}\}_{m=1}^i) \right].
\end{align}
Here $N$ is the number of time slots and $\mathbb{E}[.]$ denotes the conditional expectation over remaining energy harvests $\{E_{j,i}\}_{i=n+1}^N$ given the previous harvests as $\{E_{j,i}\}_{i=1}^n$.

Note that the dynamic program outlined by (\ref{eqn_online_bellman1}) is computationally difficult due to the dimension of the problem. However, for the case of i.i.d. energy harvests that we consider, it can be simplified by restricting to actions that only utilize current battery state. This is due to the expectation in (\ref{eqn_online_bellman1}) being independent of past energy harvests. This implies simplifying the actions in (\ref{eqn_online_action1}) as
\begin{subequations}
\label{eqn_online_action2}
\begin{align}
	p_{j,n}&=\phi_{j,n}(E_{1,n}^{bat},E_{2,n}^{bat},E_{3,n}^{bat}), \\
	 E_{j,n}^{bat}&=\sum_{i=1}^{n-1}(E_{j,i}-\tau p_{j,i})+E_{j,n},
\end{align}
\end{subequations}
where $E_{j,n}^{bat}$ is the battery state of $T_j$ at the beginning of slot $n$. The solution to (\ref{eqn_online_bellman1}) and (\ref{eqn_online_action2}) provides the optimal online power policy for finite horizon, which we compare with the offline policy in Section~\ref{sect_numerical}.

To further simplify the problem, we additionally consider the infinite horizon problem where the optimal actions are time-invariant. We formulate this as a discounted dynamic program with the Bellman equation
\begin{align}
\label{eqn_online_bellman2}
	V(\{E_{j}^{bat}&\}) = \underset{\phi_{1},\phi_{2},\phi_{3}}{\max}  R_s(\phi_j(\{E_{k}^{bat}\})) \nonumber \\
	&+ \beta \mathbb{E} \left[ V(\{E_{j}^{bat}-\tau \phi_j(\{E_{k}^{bat}\})+E_{j} \}) \right],
\end{align}
where $\phi_{j}=\phi_{j,n}$ for all $n$ and the expectation is over $E_j$, $j=1,2,3$. This equation can be solved with value iteration \cite{bertsekas1995dynamic}. Namely, starting from arbitrary initial actions, all actions $\phi_j$ are updated as the arguments that maximize (\ref{eqn_online_bellman2}), and value functions $V(\{E_{j}^{bat}\})$ are updated as in (\ref{eqn_online_bellman2}), until all actions converge to some $\phi_j^*$. Here, the {\em discount factor} $\beta < 1$ ensures that the values $V(\{E_{j}^{bat}\})$ remain bounded \cite{bertsekas1995dynamic}. The resulting actions yield an online policy that is optimal under the action restrictions in (\ref{eqn_online_action2}) and infinite transmission assumption. Hence, we refer to this policy as the optimal online policy for an infinite horizon.

\begin{remark}
Each value iteration step requires $K^3$ value updates, where $K$ is the number of transmit power values after discretization. Hence, the running time of the finite horizon algorithm is $O(NK^3)$ for $N$ epochs, and storing the optimal policy requires $O(NK^3)$ space. On the other hand, the optimal policy is time-invariant for the infinite horizon case, and requires only $O(K^3)$ space.
\end{remark}

\section{Numerical Results}
\label{sect_numerical}

In this section, we demonstrate the optimal policies for the two-way relay channel and compare the performance of the schemes in Section~\ref{sub_model_rates} and Section~\ref{sect_hybrid} in the EH setting. In simulations, energy arrivals to node $T_j$ are generated independently from a uniform distribution over $[0,E_{h,j}]$ for $j=1,2,3$, with unit epoch lengths $l_n=1$~s. The noise density is $10^{-19}$~W/Hz at all nodes and the bandwidth is $1$~MHz.

\begin{figure}
\centering
\hspace{-8pt}
\subfloat[]{\includegraphics[width=\linewidth]{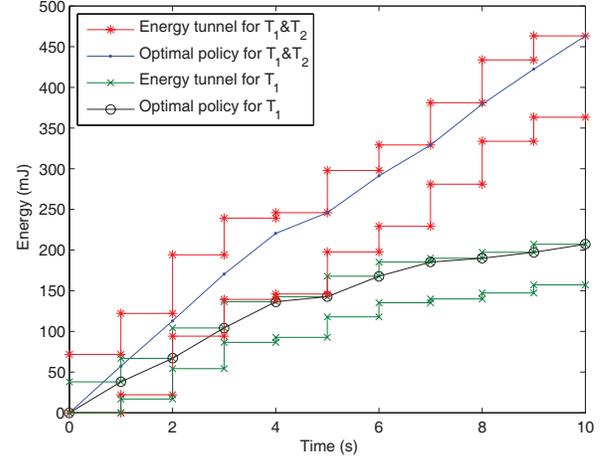}} \\  
\subfloat[]{\includegraphics[width=\linewidth]{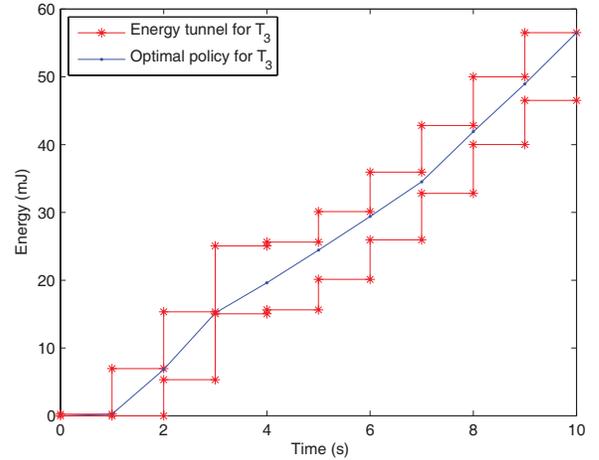}}
\caption{Optimal cumulative harvested energy and consumed energy policies for (a) node $T_1$ and sum of $T_1$ and $T_2$, and (b) node $T_3$,
for an asymmetric full-duplex channel with decode-and-forward relaying, $h_{13}=-110$~dB, $h_{23}=-116$~dB, peak energy harvesting rates
$E_{h,1}=E_{h,2}=50$~mJ and $E_{h,3}=10$~mJ, battery sizes $E_{1,max}=E_{2,max}=50$~mJ and $E_{3,max}=10$~mJ.}
\label{fig_subgradient_fd2}
\end{figure}

\begin{figure}
\centering
\subfloat[]{\includegraphics[width=\linewidth]{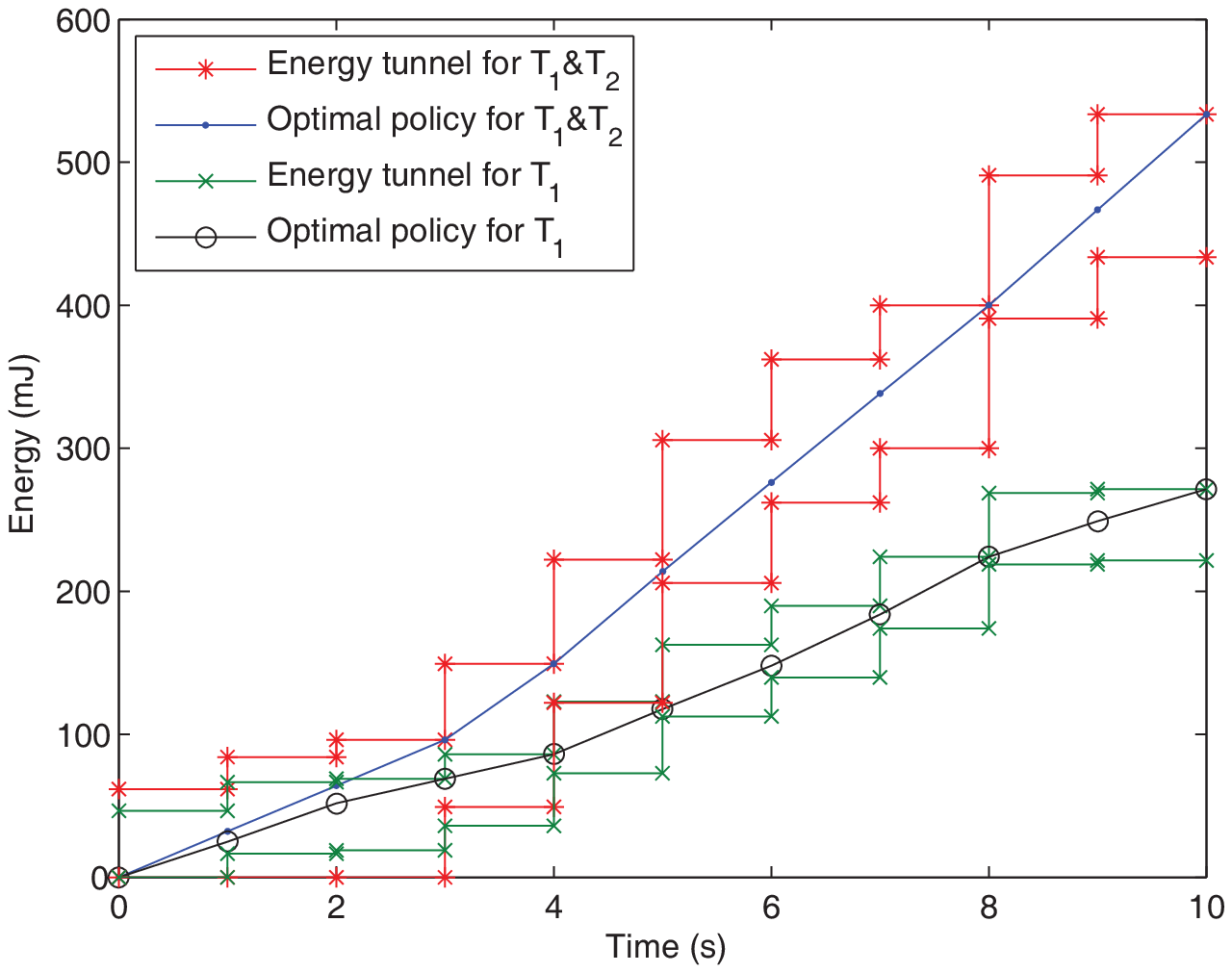}} \\ \hspace{-10pt}
\subfloat[]{\includegraphics[width=\linewidth]{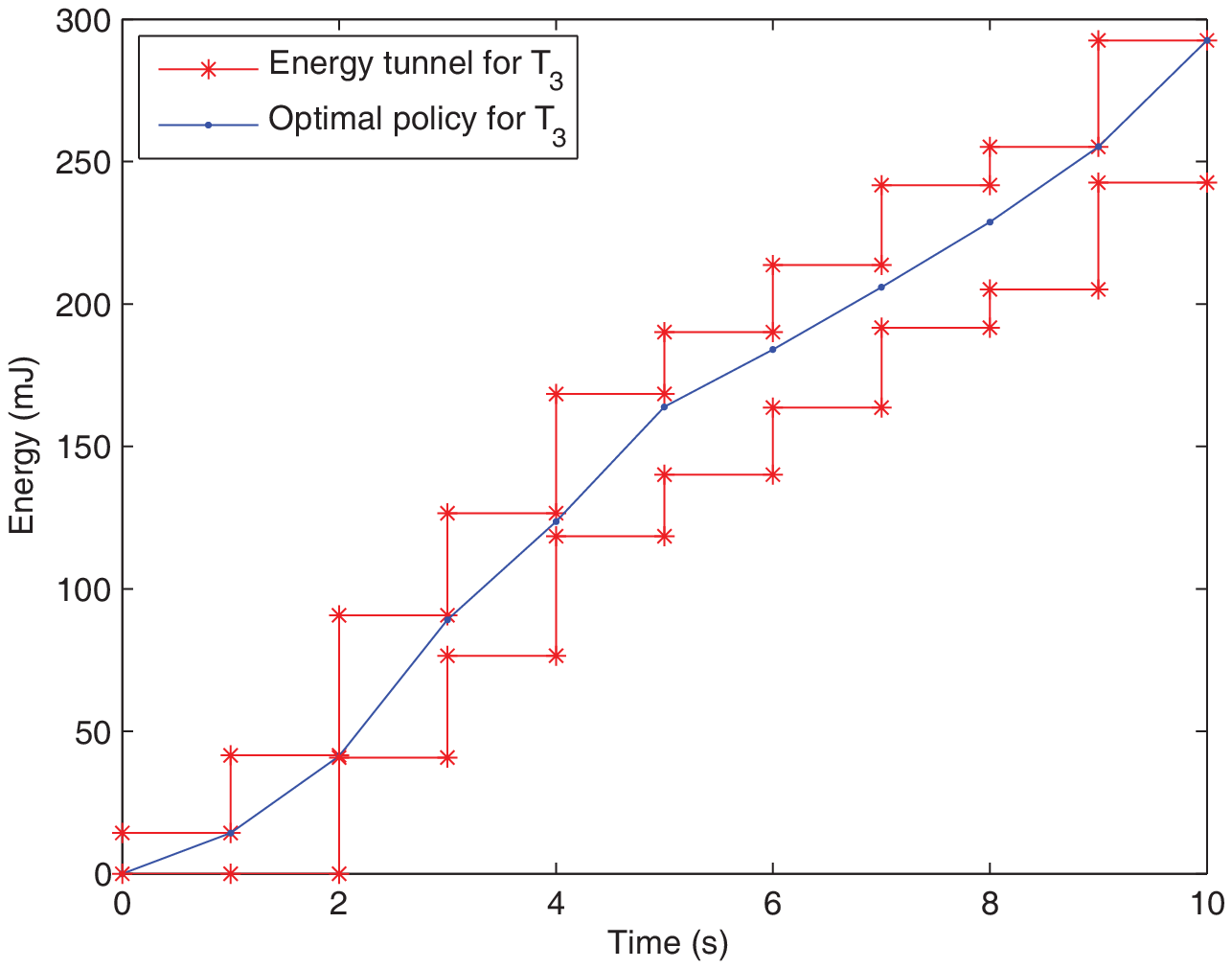}}
\caption{Optimal cumulative harvested energy and consumed energy policies for (a) node $T_1$ and sum of $T_1$ and $T_2$, and (b) node $T_3$,
for a symmetric full-duplex channel with decode-and-forward relaying and $h_{13}=h_{23}=-110$~dB, peak energy harvesting rates
$E_{h,1}=E_{h,2}=E_{h,3}=50$~mJ, battery sizes $E_{1,max}=E_{2,max}=E_{3,max}=50$~mJ.}
\label{fig_subgradient_fd1}
\end{figure}

\begin{figure}
\centering
\subfloat[]{\includegraphics[width=\linewidth]{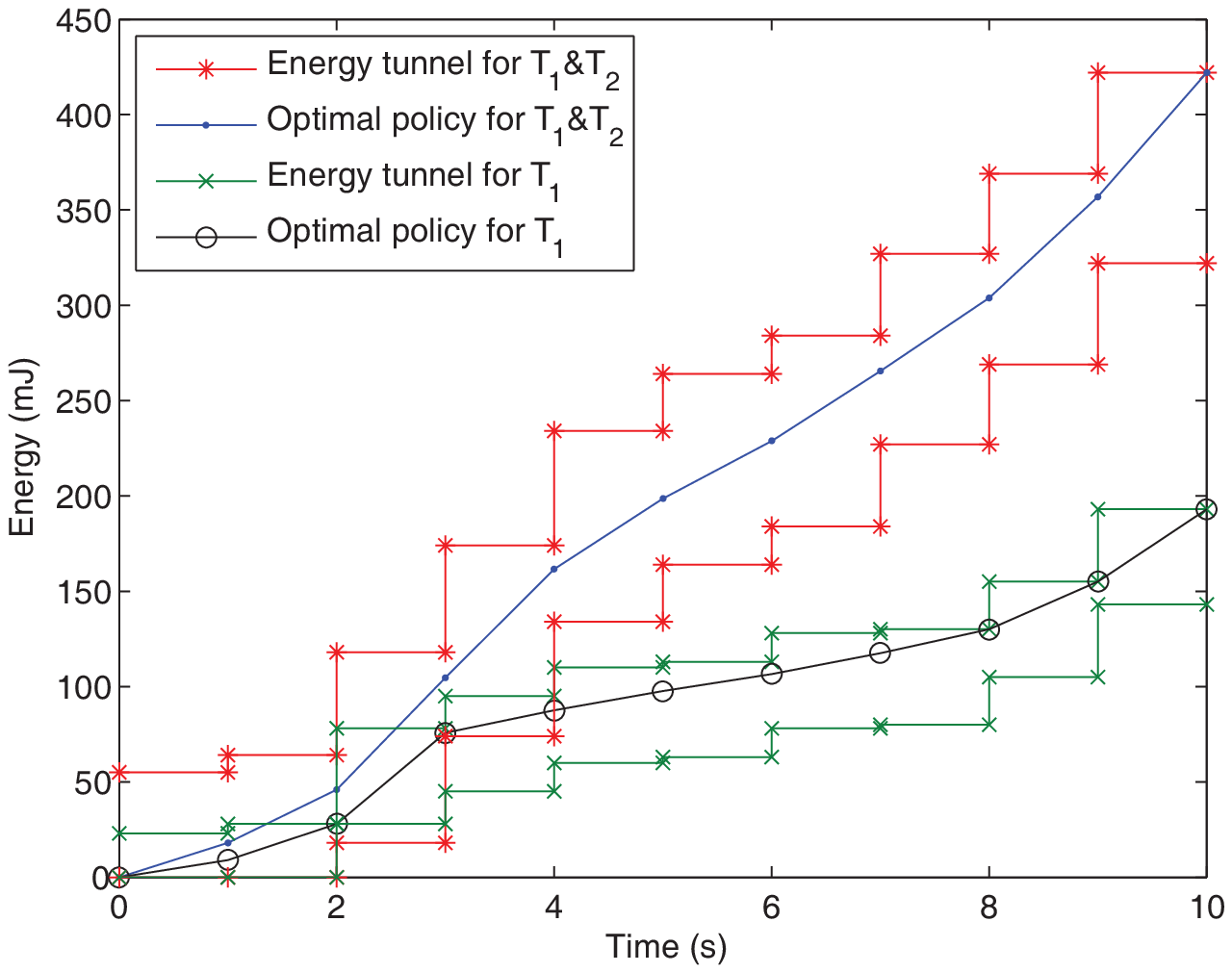}} \\
\subfloat[]{\includegraphics[width=\linewidth]{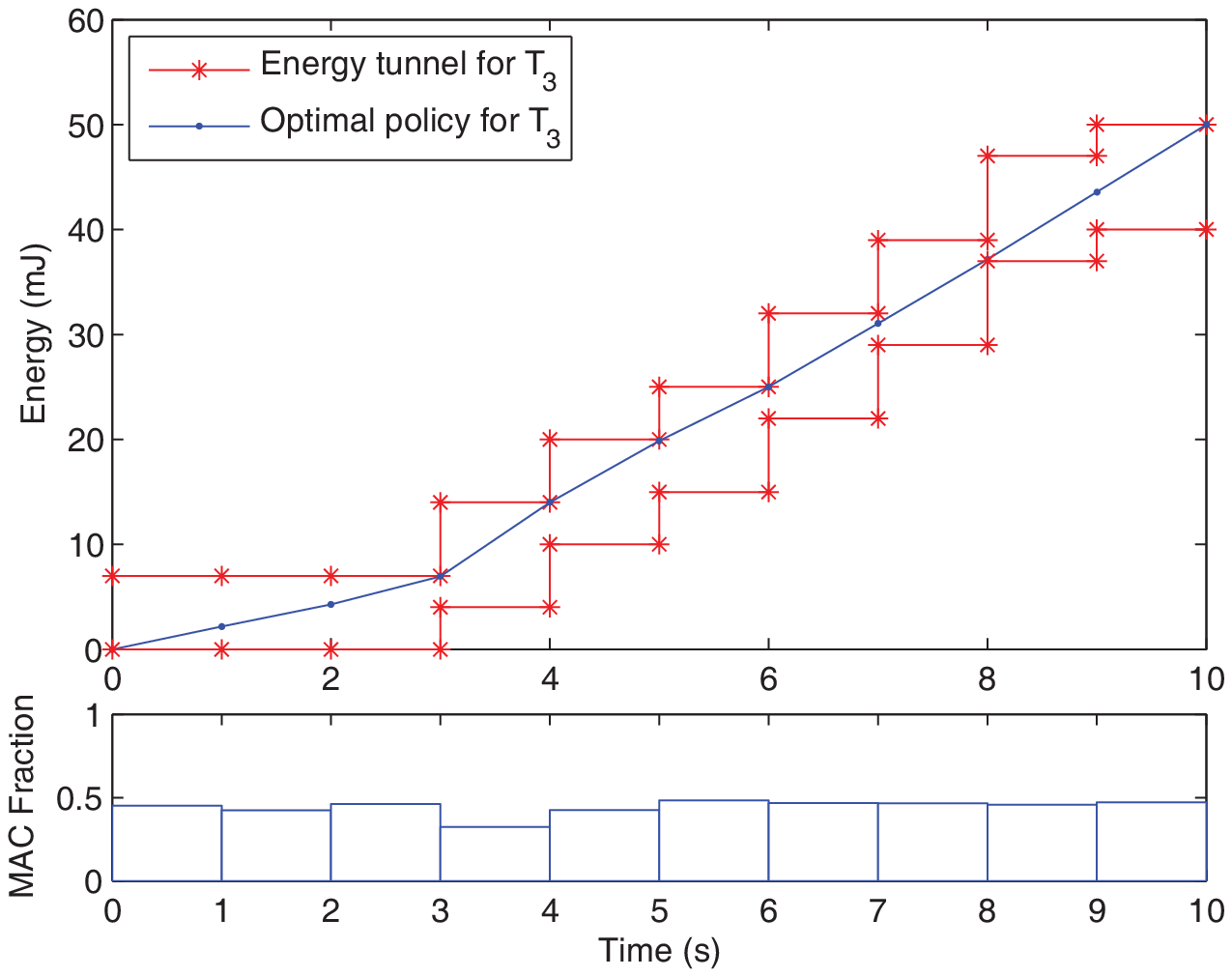}}
\caption{Optimal cumulative harvested energy and consumed energy policies for (a) node $T_1$ and sum of $T_1$ and $T_2$, and (b) node $T_3$,
for an asymmetric half-duplex channel with decode-and-forward relaying and $h_{13}=-110$~dB, $h_{23}=-116$~dB, peak energy harvesting rates
$E_{h,1}=E_{h,2}=50$~mJ, $E_{h,3}=10$~mJ and battery sizes $E_{1,max}=E_{2,max}=50$~mJ and $E_{3,max}=10$~mJ.}
\label{fig_subgradient_hd2}
\end{figure}

\begin{figure}
\centering
\subfloat[]{\includegraphics[width=\linewidth]{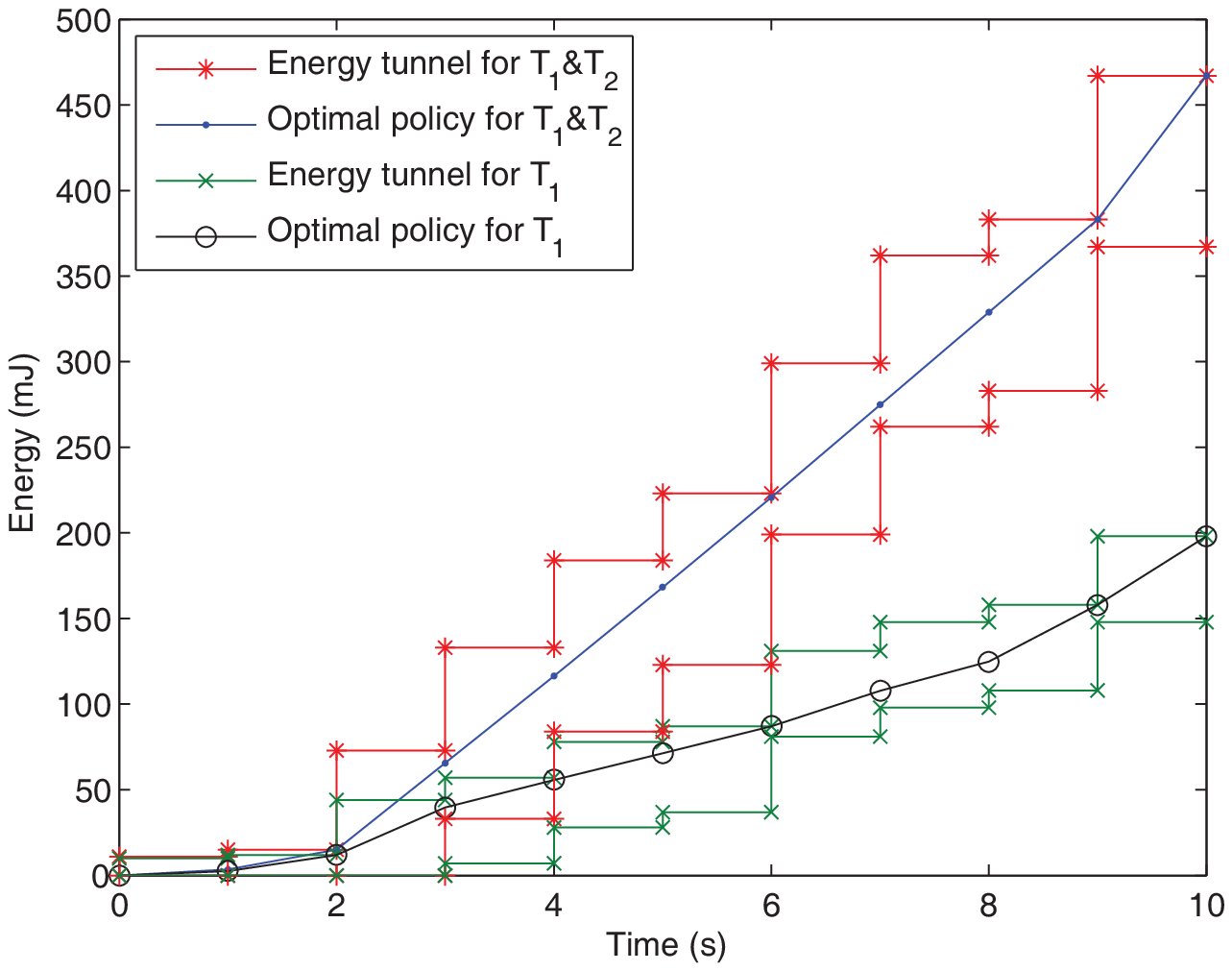}} \\
\subfloat[]{\includegraphics[width=\linewidth]{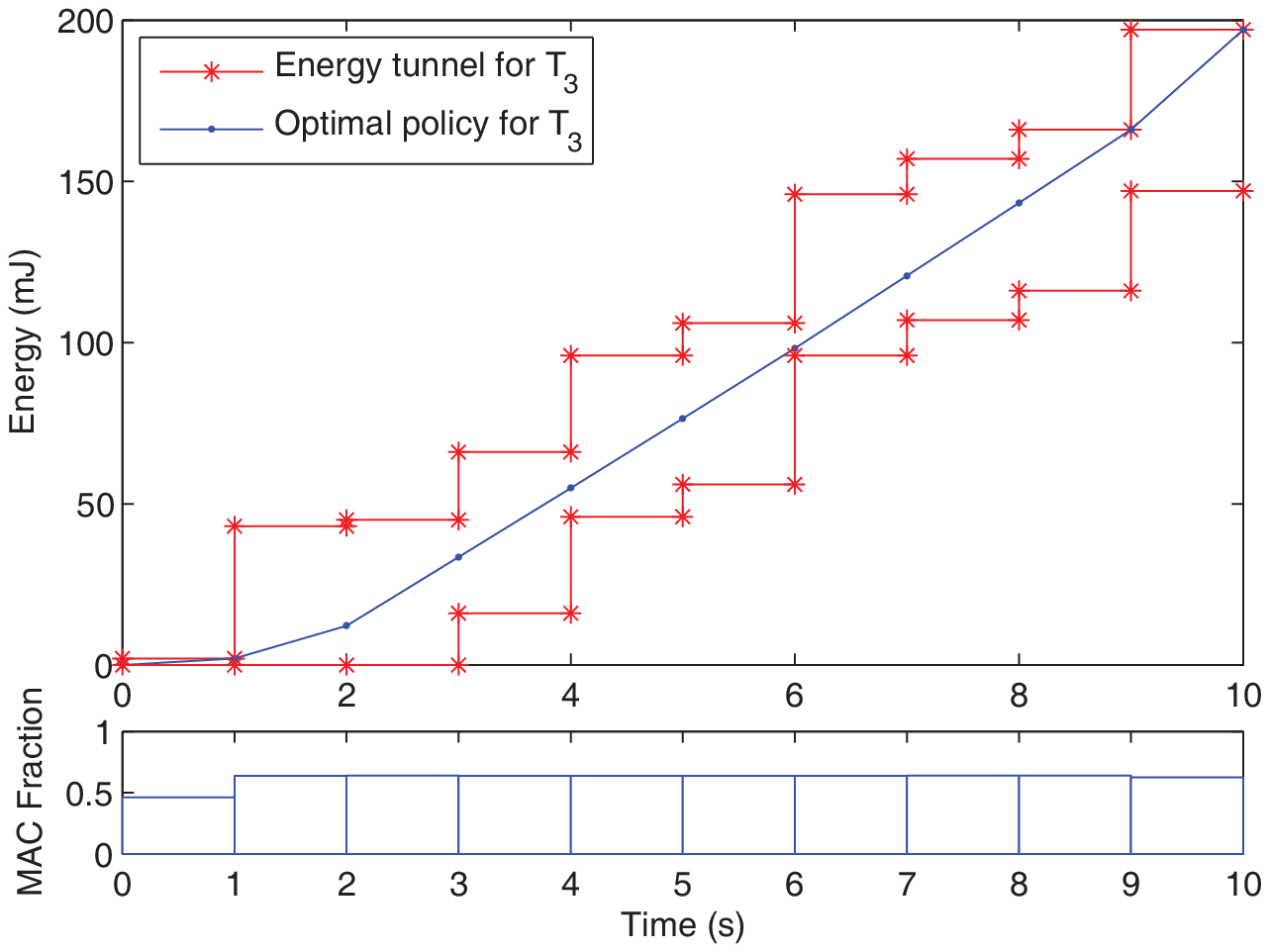}}
\caption{Optimal cumulative harvested energy and consumed energy policies for (a) node $T_1$ and sum of $T_1$ and $T_2$, and (b) node $T_3$,
for a symmetric half-duplex channel with decode-and-forward relaying and $h_{13}=h_{23}=-110$~dB, peak energy harvesting rates
$E_{h,1}=E_{h,2}=E_{h,3}=50$~mJ, battery sizes $E_{1,max}=E_{2,max}=E_{3,max}=50$~mJ.}
\label{fig_giwftunnel_df}
\end{figure}

Examples for the optimal transmit power policies found using the algorithm described in Section~\ref{sect_identify} are shown in Figures~\ref{fig_subgradient_fd2}--\ref{fig_giwftunnel_df} for decode-and-forward relaying. In each figure, cumulative energy consumed by the nodes for transmission are plotted, the derivative of which yields the average transmit powers of the nodes in each epoch. In the figures, $T_1$\&$T_2$ stands for the total cumulative energy of the nodes $T_1$ and $T_2$, and MAC fraction represents the fraction of the multiple access phase, i.e., $\Delta_n$. We remark that concavified sum-rate functions are used for the simulations, and average transmit powers are shown in the plots for clarity. Pairs of staircases, shown in red and green, represent energy causality and battery capacity constraints on the cumulative power, which is referred to as the feasible energy tunnel \cite{tutuncuoglu2012optimum}. A feasible policy remains between these two constraints throughout the transmission period. Figures~\ref{fig_subgradient_fd2} and \ref{fig_subgradient_fd1} are plotted for full-duplex nodes while Figures~\ref{fig_subgradient_hd2} and \ref{fig_giwftunnel_df} are plotted for half-duplex nodes. Both scenarios are considered for an asymmetric EH-TWRC with $h_{13} \neq h_{23}$ in Figures~\ref{fig_subgradient_fd2} and \ref{fig_subgradient_hd2}, and for a symmetric EH-TWRC with $h_{13} = h_{23}$ in Figures~\ref{fig_subgradient_fd1} and \ref{fig_giwftunnel_df}.


We remark that unlike previous work with simpler channel models, e.g., \cite{yang2012optimal, tutuncuoglu2012optimum, yang2012mac, ozel2012optimalbc}, the optimal cumulative energy or sum-power policy is not necessarily the shortest path that traverses the feasible tunnel. Figure~\ref{fig_subgradient_fd2} shows a setting with $E_{h,1}=E_{h_2}=50$~mJ and $E_{h,3}=10$~mJ, i.e., the relay is energy deprived compared to $T_1$ and $T_2$. Hence, energy efficiency is critical for the relay while this is not necessarily the case for the remaining nodes that are relatively energy-rich. This results in the optimal policy being largely dictated by the relay. Note that in Figure~\ref{fig_subgradient_fd2}, the relay follows a cumulative energy that resembles the shortest path through the feasible energy tunnel, while for $T_1$ and $T_2$ this is not the case. In contrast, in Figure~\ref{fig_subgradient_fd1}, the multiple access phase is more likely to be limiting because the sum-rate with equal transmit powers at all nodes is limited by the sum-rate constraint of the multiple access phase, see (\ref{eqn_rates_hd_mabc_sum}). Thus, the total cumulative energy, denoted with $T_1$\&$T_2$ in Figure~\ref{fig_subgradient_fd1}(a), follows the shortest path within the tunnel, similar to the optimal policy for the multiple access channel in \cite{yang2012mac}. However, broadcast powers do not yield binding constraints, implying that contrary to the energy harvesting models previously studied, e.g., \cite{yang2012optimal,tutuncuoglu2012optimum}, the optimal policy for the EH-TWRC is not necessarily unique.

Comparing Figures~\ref{fig_subgradient_hd2} and \ref{fig_giwftunnel_df}, which show optimal policies for the half-duplex model, we observe that the time division parameters $\Delta_n$ play an important role in helping energy deprived nodes. By properly selecting $\Delta_n$, the effect of unbalanced energy harvests at the sources and the relay can be mitigated. However, this still does not imply the shortest path is optimal for each node. This is due to the interplay of transmit powers though the joint rate function in the objective. Hence, whenever the transmit power changes for one user due to a full battery or an empty battery, the transmit powers of other users are affected as well. Examples to this phenomena can be found in Figure~\ref{fig_subgradient_hd2}, at $t=3,4$~s, where the energy depletion in $T_3$ is observed to affect the transmit powers of $T_1$ and $T_2$, and in Figure~\ref{fig_giwftunnel_df}, at $t=2$~s, where the energy depletion in $T_1$ and $T_2$ is observed to affect the transmit power of $T_3$. 

\begin{remark}
Similar results were observed for compress-and-forward, compute-and-forward, and amplify-and-forward relaying through simulations. We observed that identical energy harvesting profiles and channel parameters yield transmit powers that only differ slightly among relaying schemes. However, the multiple access phase fractions, $\Delta_n, n=1,2,\dots,N$, differ notably among relaying schemes in order to achieve matching multiple access and broadcast rates within each epoch. Due to the similarity of the transmit power policies, to avoid repetition, we omit the plots for these schemes.
\end{remark}

\begin{figure}[t]
\centering
\includegraphics[width=\linewidth]{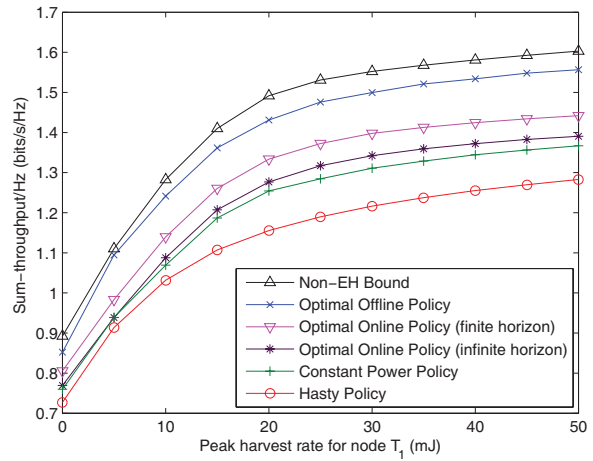}
\caption{Sum-throughput with optimal power allocations for decode-and-forward relaying compared with a non-EH upper bound, hasty policy and constant power policy.}
\label{fig_comparison}
\end{figure}

Next, we compare the performance of the optimal offline and online policies with upper and lower bounds for a decode-and-forward relay. We obtain a non-energy-harvesting upper bound by providing the total energy harvested by each node at the beginning of the transmission without a battery restriction. We also present two \naive transmit power policies, namely the hasty policy and the constant power policy, as lower bounds. The former policy, also referred to as the spend-as-you-get algorithm \cite{gorlatova2011performance}, consumes all harvested energy immediately within the same epoch. The latter policy chooses the average harvest rate at each node as the desired transmit power, and transmits with this power whenever possible. For both \naive policies, the phase fraction parameters $\Delta_n$ that maximize the instantaneous sum-rate for the given transmit powers are chosen within each epoch. We consider a half-duplex EH-TWRC with $h_{13}=h_{23}=-110$~dB, and choose the energy harvests for node $T_j$ to be independent and uniformly distributed over $[0,E_{h,j}]$ where $E_{h,2}=50$~mJ and $E_{h,3}=20$~mJ are the peak harvest rates. The infinite horizon online policy is found using a discount factor of $\beta = 0.999$. The sum-throughput values resulting from these policies with a half-duplex relay in $N=10$ epochs, averaged over $100$ independently generated scenarios, are plotted in Figure~\ref{fig_comparison}. In the figure, the peak harvest rate for node $T_1$, $E_{h,1}$, is varied in order to evaluate the performance of the policies at different harvesting rate scenarios. We observe that the optimal online policy, found for a horizon of $N = 10$ epochs, as well as its infinite horizon counterpart perform notably better than the \naive policies.

\begin{figure}[t]
\centering
\includegraphics[width=\linewidth]{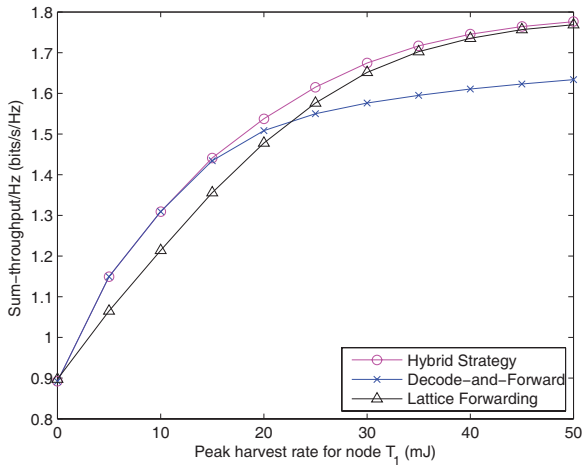}
\caption{Sum-throughput with various relaying strategies against peak harvest rates for node $T_1$. The compress-and-forward and amplify-and-forward strategies and omitted since they perform notably worse than those in the plot.}
\label{fig_sims_hybrid_rates}
\end{figure}

Finally, we compare the sum-throughput resulting from decode-and-forward, compute-and-forward, compress-and-forward, amplify-and-forward, and hybrid strategies in an EH-TWRC. The same parameters as in Figure~\ref{fig_comparison} are used in simulations. The sum-throughput values obtained over a duration of $N=10$ epochs are plotted in Figure~\ref{fig_sims_hybrid_rates}. We observe that for low and high transmit powers, either decode-and-forward or compute-and-forward outperforms the other, respectively, while they both exceed the sum-throughput values of compress-and-forward and amplify-and-forward relaying. However, as expected, the hybrid strategy outperforms all single-strategy approaches, since it performs at least as good as the best one in each epoch.

\section{Conclusion}
\label{sect_conclusion}

In this paper, we considered the sum-throughput maximization problem in a two-way relay channel where all nodes are energy harvesting with limited battery storage, i.e., finite battery. We considered decode-and-forward, compress-and-forward, compute-and-forward and amplify-and-forward relaying strategies with full-duplex and half-duplex radios. Noticing that the best relaying strategy depends on instantaneous transmit powers, we proposed a hybrid relaying scheme that switches between relaying strategies based on instantaneous transmit powers. We solved the sum-throughput maximization problem for the EH-TWRC using an iterative generalized directional water-filling algorithm. For cases where offline information about energy harvests is not available, we formulated dynamic programs which yield optimal online transmit power policies. Simulation results confirmed the benefit of the hybrid strategy over individual relaying strategies, and the improvement in sum-throughput with optimal power policies over na\"{\i}ve power policies. The online policies found via dynamic programming also proved to perform better than their na\"{\i}ve alternatives. It was observed that in a two-way channel with energy harvesting nodes, either of the communication phases, i.e., broadcast or multiple access phases, can be limiting, impacting the optimal transmit powers in the non-limiting phase as well. Thus, the jointly optimal policies were observed not to be the throughput maximizers for each individual node, or the sum-throughput maximizers for a subset of nodes - a fundamental departure in the structure of optimal policies in previous work \cite{yang2012optimal, tutuncuoglu2012optimum, ozel2011fading, tutuncuoglu2012ita, yang2012mac, yang2012broadcasting}.

We remark that the offline throughput maximization problem for the full-duplex and half-duplex cases when decode-and-forward relaying is used can also be solved using the subgradient descent algorithm as shown in \cite{tutuncuoglu2013energy}. Future directions for this channel model include optimal offline and online power policies for more involved models with data arrivals at the sources, data buffers at the relay, or a direct channel between sources. 


\end{document}